\documentclass{aa}

\newcommand{\sech}      {\mbox{$ \rm sech     $}}
\newcommand{\iso}       {\mbox{$ \rm sech^{2} $}}
\newcommand{\Rmax}      {\mbox{$  R_{\rm max} $}}
\newcommand{\zo}        {\mbox{$  z_{0}       $}}

\newcommand{\mum}	{\mbox{$ \mu \rm m    $}}

\newcommand{\A}         {\mbox{$ ^{\rm a)}    $}}

\usepackage{graphicx}
\graphicspath{{/home/schwarz/bochum/papers/aa/99_1_print}{/ps/}}

\begin{document}

   \thesaurus{20         
              (11.07.1;  
               11.09.2;  
               11.16.1;  
               11.19.2;  
               11.19.7;  
               13.09.1)} 

   \title{The influence of interactions and minor mergers \\
          on the structure of galactic disks
          \thanks{Based on observations obtained at the European Southern Observatory
          (ESO, La Silla, Chile), Calar Alto Observatory operated by the MPIA (DSAZ, Spain),
          Lowell Observatory (Flagstaff/AZ, USA), and Hoher List Observatory (Germany).}
         }

   \subtitle{I. Observations and disk models}

   \authorrunning{U. Schwarzkopf \& R.-J. Dettmar}
   \titlerunning{The influence of interactions and minor mergers on disk structure. I}

   \author{U. Schwarzkopf \hspace{-0.8mm} \inst{1,}\inst{2,}\thanks{\emph{Present address:}
           Steward Observatory, University of Arizona,
           933 N. Cherry Ave., Tucson, Arizona 85721, USA}
          \and
           R.-J. Dettmar \inst{1}}

   \offprints{U. Schwarzkopf \\
              (schwarz@as.arizona.edu)}

   \institute{Astronomisches Institut, Ruhr-Universit\"at Bochum,
              Universit\"atsstra{\ss}e 150, 44780 Bochum, Germany
             \and
              Steward Observatory, University of Arizona,
              933 N. Cherry Ave., Tucson, Arizona 85721, USA}

   \date{Received 22 July 1999 / Accepted 6 March 2000}

   \maketitle

   \begin{abstract}

This paper is the first part in our series on the influence of tidal interactions and
minor mergers on the radial and vertical disk structure of spiral galaxies. We report
on the sample selection, our observations, and data reduction. Surface photometry of
the optical and near infrared data of a sample of 110 highly-inclined/edge-on disk
galaxies are presented. This sample consists of two subsamples of 61 non-interacting
galaxies (control sample) and of 49 interacting galaxies/minor merging candidates.
Additionally, 41 of these galaxies were observed in the near infrared.
We show that the distribution of morphological types of both subsamples is almost
indistinguishable, covering the range between $0 \le T \le 9$. An improved, 3-dimensional
disk modelling- and fitting procedure is described in order to analyze and to compare
the disk structure of our sample galaxies by using characteristic parameters. We find
that the vertical brightness profiles of galactic disks respond very sensitive even to
small deviations from the perfect edge-on orientation. Hence, projection effects of
slightly inclined disks may cause substantial changes in the value of the disk scale
height and must therefore be considered in the subsequent study.


   \end{abstract}
%

\section{Introduction}

\noindent
Considering the fact that the majority of (spiral) galaxies is not completely isolated
but located in an environment which enables repeated close encounters or even merging
processes with small companions it seems to be meaningful to systematically investigate
the properties of galaxies affected by such processes. The investigation of their
structural and dynamical changes caused by tidal interactions or low-mass satellite infall --
hence ``minor merger'' -- can help to clarify how far the evolution of disk galaxies was
modified or even dominated by environmental effects.

\medskip

Several N-body simulations were performed during the last decade in order
to study the influence of minor mergers on galactic disks in greater detail
(e.g. Quinn et al. \cite{quinn1993}; Mihos et al. \cite{mihos1995}; Walker
et al. \cite{walker1996}). It was possible to use more realistic, multiple-component models
for the galaxy-satellite system -- usually consisting of disk, bulge, and halo -- as well
as a large number of particles ($n_{\rm disk} \geq 32 \, 000$). One of the main
conclusions was that even merging processes in the range between $M_{\rm sat}/M_{\rm disk}
\approx 0.05 - 0.2$ can cause a vertical thickening of the stellar disk component by a
factor between 2 and 4, depending on the galactocentric distance. It was found that this
vertical heating is due to a gain of kinetic energy of the disk stars by enhanced two-body
relaxation. According to a series of papers on the frequency of these so called ``soft merging''
events (e.g. Toth \& Ostriker \cite{toth1992}; Zaritsky \cite{zaritsky1995}, \cite{zaritsky1996})
a large number of present-day (disk-) galaxies were affected by merging- or accretion processes
of this magnitude since they have formed. As a consequence, interactions and minor mergers
within this mass range might modify our picture of galaxy formation and evolution.

However, the enormous parameter space of such a complex scenario makes it difficult to derive
general conclusions from a set of few specific simulations. The quantitative results still
crucially depend on the chosen parameters such as the content and behaviour of gas in the disk,
the mass ratio between bulge and disk, induced star formation, or the satellite orbit
(Quinn et al. \cite{quinn1993}; Mihos et al. \cite{mihos1995}; Velazquez \& White
\cite{velazquez1999}).

\medskip

Statistical studies of galaxy interactions -- based on optical photometry of disk
galaxies (Reshetnikov \& Combes \cite{reshetnikov1996}, \cite{reshetnikov1997}) -- focused
on the effects of tidally-triggered disk thickening between systems of comparable mass.
They found that the ratio $h/\zo$ of the radial exponential scale length $h$ to
the constant scale height $\zo$ is only about twice smaller for interacting galaxies -- a
lower value than derived from the minor mergers simulations. However, the small number of
objects in their sample (7 non-interacting and 24 interacting galaxies) did not permit to
study these questions in detail.

Therefore, we started a project based on a larger sample of edge-on disk galaxies in both
optical and near infrared passbands. This combination offers a number of advantages:

\noindent
First, observations in the near infrared particularly benefit from the much lower
dust extinction near the galactic plane, i.e. at small $z$.
Second, the presence of a dust lane along the major axis of most edge-on disk galaxies
still presents one of the best methods to determine precise inclinations of the disks --
two facts that will become very important in order to derive reliable scale parameters
from a disk fitting procedure. Third, this combination enables us to make conclusions
on disk populations of different ages.

\smallskip

The main questions of this study can be summarized as follows:

\begin{itemize}

\vspace{-2mm}

\item Are interactions/minor mergers able to change the radial and vertical structure
of affected galactic disks?

\smallskip

\item Is there a substantial vertical disk thickening?

\smallskip

\item Of which order are the differences and similarities in the disk parameter
distribution for a sample of interacting/non-interacting galaxies, respectively?

\smallskip

\item To what extent are the disk properties of galaxies in the local universe
influenced by interactions/minor mergers?

\end{itemize}

\noindent
Due to the complexity of these questions the paper is split into three parts:
in this first part (Paper~I) we present deep optical and near infrared photometric
data of a total sample of 110 highly-inclined/edge-on disk galaxies. This sample
consists of two subsamples of 61 non-interacting galaxies (control sample) and of 49
minor merging candidates. Additionally, 41 of these galaxies were observed in the near
infrared. In Sect.~2 the criteria of the sample selection will be described briefly.
Sect.~3 gives an overview on the observations and data reduction. The disk modelling-
and fitting procedure applied to derive the disk parameters will be reviewed in Sect.~4.
In Sect.~5 we summarize and conclude the paper.

In the second part (Schwarzkopf \& Dettmar \cite{schwarzkopf2000_II}, Paper~II) the
results of a detailed analysis of the structure of galactic disks will be presented.

The third part (Schwarzkopf \& Dettmar in preparation, Paper~III) will be focused on
the influence of accompanying minor merger features -- like disk ``warping'' and
``flaring'' -- on the vertical disk structure.




%
%
%

\tabcolsep1.55mm

\begin{table*}[t]
  \caption[ ]{Nomenclature of morphological galaxy types (cross reference). \\
   columns: (1) Source: ESO-LV= Lauberts \& Valentijn (\cite{lauberts1989}),
   Hubble= Hubble (\cite{hubble1926}), \\
   ESO-Upp= ``old'' type in Lauberts \& Valentijn (\cite{lauberts1989});
   (2) Designation of morphological types.}
  \label{types}
  \begin{flushleft}
  \begin{tabular}{llccccccccccccccccc}
  \cline{1-19}
  \hline\hline\noalign{\smallskip}
  \multicolumn{1}{l}{\bf Source}   && \multicolumn{17}{c}{\bf Morphological Type}  \\
  \noalign{\smallskip}
  \multicolumn{1}{c}{(1)} && \multicolumn{17}{c}{(2)} \\
  \noalign{\smallskip}
  \hline\noalign{\medskip}
  \multicolumn{1}{l}{ESO-LV}  && -5 & -4 &  -3  & -2 & -1 &  0   & 1  &  2   & 3  &  4   & 5    &
   6  & 7  &   8    &  9 &  10  & 11 \\
  \noalign{\smallskip}
  \cline{1-19}
  \noalign{\smallskip}
  \multicolumn{1}{l}{Hubble}  &&  E & -- & E-S0 & S0 & -- & S0/a & Sa & Sa-b & Sb & Sb-c & --   &
   Sc & -- & Sc-Irr & -- & IrrI & -- \\
  \noalign{\smallskip}
  \multicolumn{1}{l}{ESO-Upp} &&  E & -- & E-S0 & S0 & -- & S0-a & Sa & Sa-b & Sb & Sb-c & S... &
   Sc, Sc-d & S.../Irr & Sd & -- & Irr & unknown \\
  \noalign{\smallskip}
  \hline
  \end{tabular}
  \end{flushleft}


\vspace{69mm}

\begin{minipage}[b]{8.8cm}
\begin{picture}(8.8,7.7)
{\includegraphics[angle=90,viewport=320 430 570 720,clip,width=80mm]{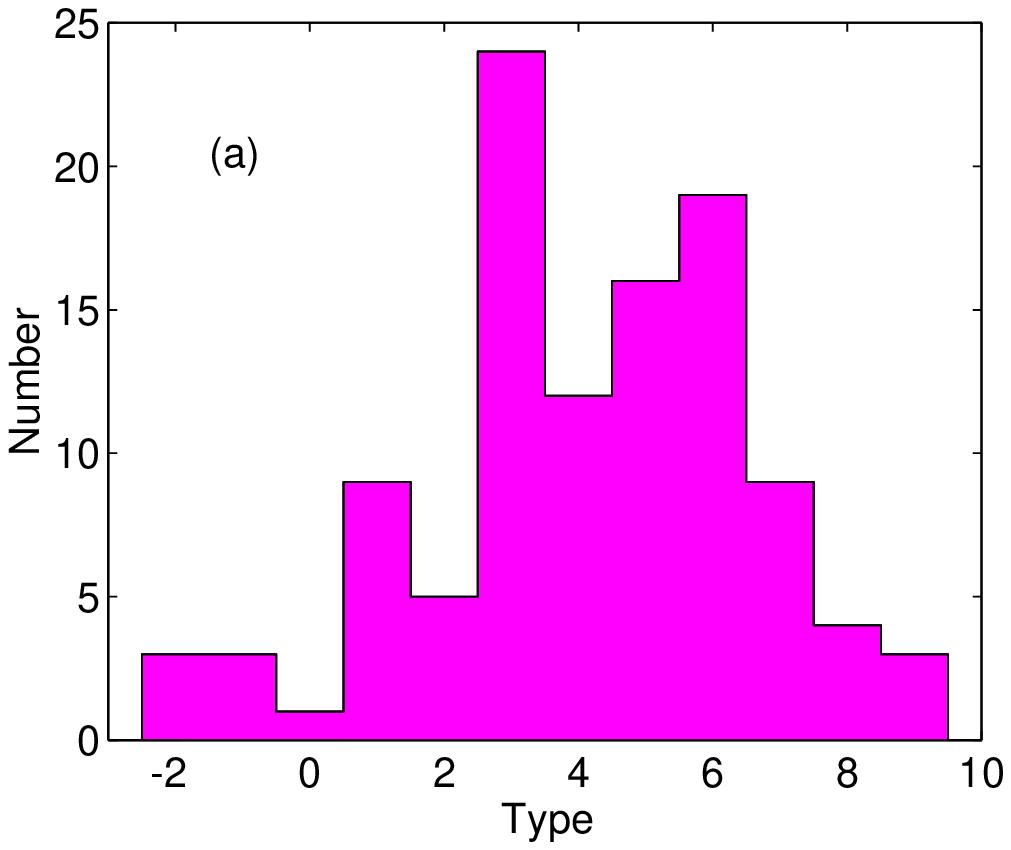}}
\end{picture}
\end{minipage}
\hfill
\begin{minipage}[b]{8.8cm}
\begin{picture}(8.8,7.7)
{\includegraphics[angle=90,viewport=320 430 570 720,clip,width=80mm]{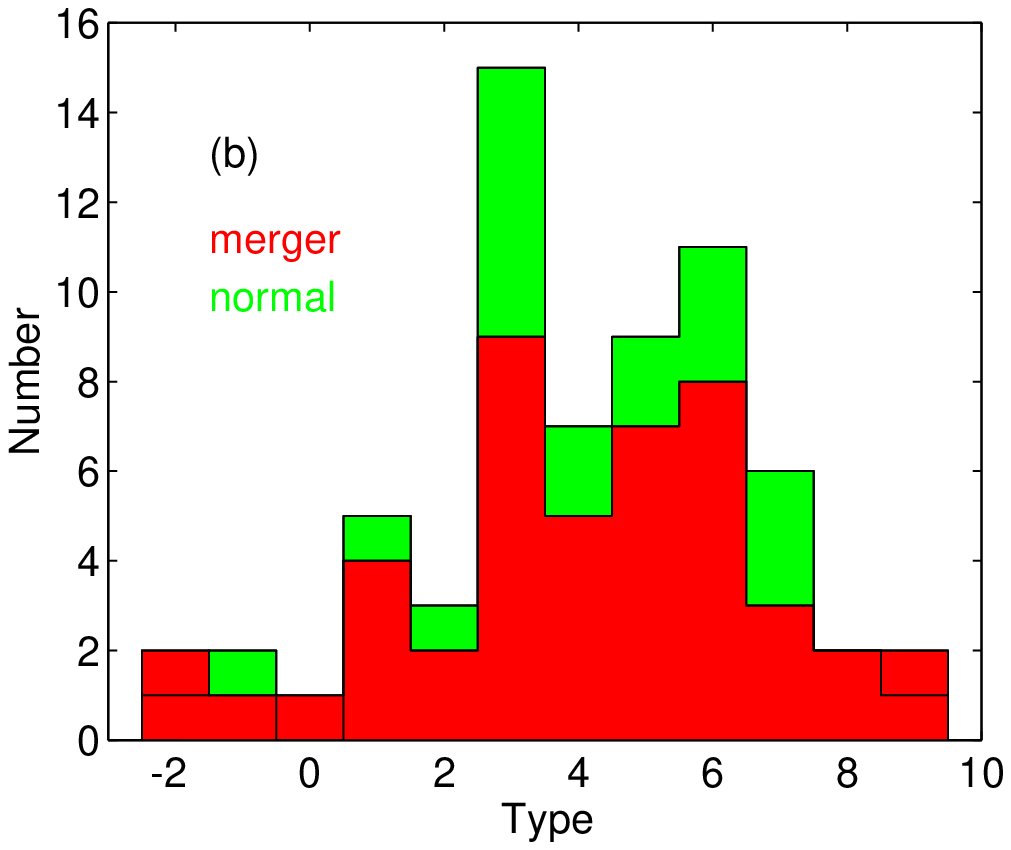}}
\end{picture}
\end{minipage}
{\bf \noindent Fig.~1a and b.} The distribution of morphological types
{\bf (a)} for the total galaxy sample and {\bf (b)} for both subsamples of
non-interacting (normal) and interacting/merging galaxies (merger).

\end{table*}


\section{Sample selection}

\subsection{General remarks}

\noindent
The selection of two separate subsamples of highly-inclined disk galaxies -- consisting
of a large number of interacting/merging candidates in the mass range
$M_{\rm sat}/M_{\rm disk} \approx 0.1 $, and of relatively isolated disk galaxies -- is
a crucial point for the comparison of disk parameters. Since presumably not all of the
merging processes of this order of magnitude -- compared to large merging events -- are
able to change the structure of affected galaxy disks completely, two facts must be
considered in order to classify both subsamples:

First, only merging events in a more progressed phase will have any appreciable effect
on the disk structure. We therefore did not consider candidates in an early stage of
interaction. Instead, the sample contains some highly-inclined disk galaxies that show no
separate satellite merging with the disk component, but strong evidence for accretion in
the recent past (indicated by both a disturbed disk structure and characteristic effects
such as warping and tidal tails). Most of these objects are located in galaxy groups with
a few members, e.g. NGC 3628, NGC 4634, or NGC 4762. Due to their perturbed structure a
morphological classification is difficult.

Second, a control sample of non-interacting edge-on disk galaxies must include all morphological
types in the range $0 \le T \le 9$ (Table~\ref{types}). The galaxies should be isolated without
indication of interaction and accretion. In particular, it is important to include the
latest type spirals like Sd -- so called ``superthin'' galaxies -- since they belong to
a class of disk galaxies having the smallest known axis ratios (between 1:9 and 1:20).
These galaxies are characterized by velocity curves of modest gradient
(Goad \& Roberts \cite{goad1981};
Griv \& Peter \cite{griv1996a},\cite{griv1996b},\cite{griv1996c}),
indicating that the ``superthins'' represent kinematically almost unheated galaxy disks.

Beyond these ``limitations'' and the specific criteria described in the next section. it seems
difficult to give a generalized definition of the selection criteria. As a consequence, it
is unavoidable that the sample of non-interacting galaxies is slightly ``polluted'' by some
galaxies that were affected by interactions/minor mergers during their past but show no
indication of these effects today. This effect would, however, only influence the
non-interacting galaxies and lead to an underestimation of the actual differences
between both samples.


\subsection{The sample of interacting/merging galaxies}

\noindent
In this study, the notation ``interacting/merging'' refers closely to the classification
scheme introduced by Arp \& Madore (\cite{arp1987}) and will therefore be used as a synonym
for all galaxies which fulfil the following criteria: \\
Galaxies with interacting companions, interacting doubles, galaxies with peculiar disks, galaxies
with tails, loops of material or debris, irregular or disturbed galaxies, chains and groups of
galaxies. A more detailed description of these classifications is given in
Arp \& Madore (\cite{arp1987}).

With these limitations in mind, we selected interacting/merging candidates with an inclination
$i \ge 85 \degr$ (derived from a first visual inspection). During this pre-selection we did not
impose any morphological restrictions except that their type should be not much earlier than
$T \approx 0$. The candidates were chosen from optical prints in ``A Catalogue of Southern
Peculiar Galaxies and Associations'' (Arp \& Madore \cite{arp1987}), ``Atlas of Peculiar
Galaxies'' (Arp \cite{arp1966}), and the NASA ``Atlas of Galaxies'' (Sandage \& Bedke
\cite{sandage1988}). We also selected some systems from the catalogue ``Satellites of
Spiral Galaxies'' (Zaritsky et al. \cite{zaritsky1993}, \cite{zaritsky1997}).

Another selection criterion for the minor mergers and those galaxies in the subsample with very
close companions was the mass ratio between the main bodies. For all the candidates which could
be separated into two individuals (i.e. for $\approx 20$ galaxies of the interacting/merging
sample) this ratio was checked by an estimation of their total fluxes within a certain aperture
or ellipse that contains all intensities down to the sky brightness.
The resulting mean ratio is $M_{\rm sat}/M_{\rm disk} \approx 0.08 \pm 0.035$. Typical examples
(with errors of $\approx 0.005$) are NGC 1531/32 ($\approx 0.05$), NGC 128 ($\approx 0.045$)
or NGC 1888 ($\approx 0.07$). For interacting galaxies located a less dense group or for
those with a remote companion this mass ratio can be larger.


\subsection{The sample of non-interacting galaxies}

\noindent
The selection criteria for disk inclination and morphological types were the same as described
for the interacting/merging galaxies.
Our principal sources for the subsample of highly-inclined, non-interacting galaxies were the
ESO-Uppsala catalogue (Lauberts \& Valentijn \cite{lauberts1989}), the RC3- (de Vaucouleurs
et al. \cite{vaucouleurs1991}) and UGC (Nilson \cite{nilson1973}) catalogues, the ``Carnegie
Atlas of Galaxies'' (Sandage \& Bedke \cite{sandage1994}), and ``The Hubble Atlas of Galaxies''
(Sandage \cite{sandage1961}). To check their isolation, larger fields $(\approx 10' - 15'$,
depending on the distance of individual objects) were inspected visually using the Digitized
Sky Survey \footnote{The Digitized Sky Survey was produced at the Space Telescope Science
Institute under U.S. Government grant NAG W-2166.}.

In order to benefit from a better spatial resolution of some closer objects there was only
a lower limit in apparent angular disk diameter of $\ge 2 \arcmin$ for both galaxy samples.
Finally, the sample of non-interacting galaxies was filled up by 11 edge-on galaxies
of the Barteldrees \& Dettmar (\cite{barteldrees1994}) data set.


\subsection{Distribution of morphological types}

\noindent
Several studies of the properties of edge-on galaxies argued that some of the disk
parameters of spiral galaxies, e.g. the ratio of disk scale length to scale height,
might be correlated with the morphological type (de Grijs \& van der Kruit \cite{grijs1996};
de Grijs \cite{grijs1997}; Schwarzkopf \& Dettmar \cite{schwarzkopf1998}).
Therefore, it was also important to ensure that both subsamples investigated in this study
are not affected by selection effects. This applies, in particular, to the distribution of
morphologigal types and to the redshifts/distances of selected objects. Due to their relation
to the absolute properties of galactic disks the latter will be discussed in detail in
Paper~II.

The distribution of morphological types (according to NED
\footnote{The NASA/IPAC Extragalactic Database (NED) is operated by the Jet Propulsion
Laboratory, California Institute of Technology, under contract with the National
Aeronautics and Space Administration.})
is shown in Fig.~1a for the complete galaxy sample, and in Fig.~1b for both
subsamples of non-interacting and interacting/merging galaxies, respectively.
The statistical test of Kolmogorov \& Smirnov (Darling \cite{darling1957}; Sachs
\cite{sachs1992}) -- hereafter KS -- quantifies the similarity of both subsamples
with a result of 0.04. That is significantly lower than the value necessary for
the 20\%-limit (0.2), which is the strongest of the KS-criteria.
Considering also the unavoidable errors ($\Delta T \approx \pm 1$) introduced by an automated
type classification (Corwin et al. \cite{corwin1985}; Lauberts \& Valentijn \cite{lauberts1989})
both distributions are statistically indistinguishable.
The gross of galaxies covers the range between $3 \le T \le 7$, with a strong peak at $T=3$.
This peak can also be observed in the frequency distribution of galaxy types in the available
catalogues. The effect is caused by the selection criteria of the classification programs used
in the catalogues (Sect.~6 in Lauberts \& Valentijn (1989)).




\section{Observations and data reduction}

\subsection{Optical observations}

\noindent
Due to the large number of galaxies needed for reliable statistics the optical
observations were obtained with different telescopes and during several observing
runs between February 1996 and June 1998. The following telescopes were used:

1.54m Danish and 61cm Bochum telescope on La Silla, Chile;
42-inch telescope at Lowell Observatory, Flagstaff;
1.23m telescope at Calar Alto Observatory, Spain;
1.06m telescope at Hoher List Observatory, Germany.
The used passbands were Johnson $R$ and Thuan \& Gunn $r$, with a central
wavelength at $\lambda_{\rm c} = 652$ nm and 670 nm, bandpass
$\Delta \lambda = 162$ nm and 103 nm, respectively.

All observing runs and the used telescope/detector characteristics are listed in
Table~\ref{telescopes}. The seeing conditions are given as averaged FWHM values.

Details of the optical observations are listed in Table~\ref{samples}. Additionally
for most of the sample galaxies the total blue surface brightness, $B_{\rm T}$, and
the morphological type $T$ (revised Hubble type, according to Lauberts \& Valentijn
(\cite{lauberts1989}), Table~\ref{types}) are given.


\subsection{Near infrared observations}

\noindent
The near infrared observations were obtained with the MAGIC camera of the MPIA attached
to the 2.2m and 1.23m telescopes of the Calar Alto Observatory, Spain, and with the
IRAC2b camera on the ESO/MPIA 2.2m telescope of the European Southern Observatory
(ESO), Chile, respectively. Both the MAGIC and IRAC2b cameras are equipped with a Rockwell
$256 \times 256$ $\rm pixel^{2}$ NICMOS3 HgCdTe array. The data were acquired
during several observing runs between February 1996 and June 1998.

For all runs we used the $K$ filter (central wavelength  $\lambda_{\rm c} = 2.20 \mum$,
bandpass $\Delta \lambda = 0.40 \mum$), or the $K'$ filter ($\lambda_{\rm c} = 2.15 \mum,
\Delta \lambda = 0.32 \mum $). For some objects images in the $H$ filter were obtained
($\lambda_{\rm c} = 1.65 \mum, \Delta \lambda = 0.30 \mum $).


We recorded object and sky frames alternately, with typical single integration times
of $10 \times 5$s per sequence, and spatially separated by $\approx 6'$.
To eliminate bad pixels, the telescope ``on source''-positions were dithered by a few
arcseconds between subsequent exposures.
Since many of the sample galaxies are faint objects it was required to reach at least
$30 - 40$ min integration time ``on source'', which corresponds to $\approx 9 - 12$ cycle
repetitions and a resulting observing time of $\approx 2.5 - 3$ hours. Therefore, only
one third of our total optical galaxy sample was observed in the near infrared.

Since our study is aiming at an investigation of the structure and geometrical properties
of galactic disks, we could make use of observations obtained under non-photometric conditions.
For that reason the optical and near infrared contour maps of the sample galaxies shown in
Figs.~3 and 4 are not flux calibrated.

As for the optical observations, the telescope/detector characteristics as well as a list
of all galaxies observed in the near infrared are given in Tables~\ref{telescopes}
and \ref{samples}, respectively.

%
%
%

\tabcolsep0.8mm

  \begin{table}[t]
  \caption[ ]
  {Observing dates and telescope/detector characteristics, given for
   optical and near infrared observations. \\
   columns: (1) Observing dates; (2) Telescopes used: 2.2 CA=2.2m Calar Alto,
   2.2 ESO=2.2m ESO/MPIA, 1.5 DA=1.54m Danish, 1.2 CA=1.23m Calar Alto,
   1.1 LO=1.1m Lowell, 1.0 HL=1.06m Hoher List, 0.6 BO=0.6m Bochum;
   (3) Field of view; (4) Pixel scale; (5) Detector used;
   (6) Average seeing conditions (FWHM).}
  \label{telescopes}
  \medskip
  \begin{flushleft}
  \begin{tabular}{llclrcllc}
  \cline{1-9}
  \hline\hline\noalign{\smallskip}
   \multicolumn{2}{c}{Observing}          && \multicolumn{1}{c}{Telescope} &
   \multicolumn{1}{c}{Field}              && \multicolumn{1}{c}{Scale}     &
   \multicolumn{1}{c}{Detector}            & \multicolumn{1}{c}{Seeing}   \\
   \multicolumn{2}{c}{dates}          &&   & \multicolumn{1}{c}{[$'$]}    &&
   \multicolumn{1}{c}{$[''/\rm pix]$}  &   & \multicolumn{1}{c}{$['']$}   \\
   \noalign{\smallskip}
   \multicolumn{2}{c}{(1)} && \multicolumn{1}{c}{(2)} & \multicolumn{1}{c}{(3)} &&
   \multicolumn{1}{c}{(4)} &  \multicolumn{1}{c}{(5)} & \multicolumn{1}{c}{(6)} \\
  \noalign{\smallskip}
  \hline\noalign{\smallskip}
  \multicolumn{9}{c}{\bf Optical} \\
  \noalign{\smallskip}
  \hline\noalign{\smallskip}
   Mar & 1996  &&  1.1 LO      &    4.9     &&  0.729$\A$  &  TI 800       &  2.0  \\
   May & 1996  &&  1.0 HL      &   27.0     &&  0.825      &  LORAL 2048   &  3.0  \\
   Jun & 1996  &&  1.2 CA      &    8.6     &&  0.503      &  TEK 1024     &  2.0  \\
   Aug & 1996  &&  1.0 HL      &   27.0     &&  0.825      &  LORAL 2048   &  3.0  \\
   Sep & 1996  &&  1.2 CA      &    8.6     &&  0.503      &  TEK 1024     &  2.0  \\
   Dec & 1996  &&  0.6 BO      &    4.8     &&  0.496      &  TH 7882      &  1.8  \\
   Jan & 1997  &&  1.2 CA      &    8.6     &&  0.503      &  TEK 1024     &  2.0  \\
   Feb & 1997  &&  1.0 HL      &    6.8     &&  0.800$\A$  &  LORAL 2048   &  3.2  \\
   Apr & 1997  &&  1.5 DA      &   13.7     &&  0.390      &  LORAL 2048   &  1.5  \\
   May & 1997  &&  1.1 LO      &    9.9     &&  0.730$\A$  &  SITe 2k      &  2.0  \\
   Sep & 1997  &&  1.0 HL      &   27.0     &&  0.825      &  LORAL 2048   &  3.0  \\
   Jun & 1998  &&  1.5 DA      &   13.7     &&  0.390      &  LORAL 2048   &  1.6  \\
  \noalign{\smallskip}
  \hline\noalign{\smallskip}
\multicolumn{9}{c}{\bf Near Infrared} \\
  \noalign{\smallskip}
  \hline\noalign{\smallskip}
   Mar & 1996  &&  1.2 CA      &    5.1     &&  1.200      &  NICMOS3      &  1.8  \\
   Sep & 1996  &&  2.2 CA      &    2.7     &&  0.642      &  NICMOS3      &  1.8  \\
   Jan & 1997  &&  2.2 CA      &    2.7     &&  0.642      &  NICMOS3      &  1.5  \\
   Apr & 1997  &&  2.2 ESO     &    3.0     &&  0.708      &  NICMOS3      &  1.6  \\
   Feb & 1998  &&  2.2 CA      &    2.7     &&  0.642      &  NICMOS3      &  1.8  \\
   May & 1998  &&  2.2 ESO     &    3.0     &&  0.708      &  NICMOS3      &  1.5  \\
  \noalign{\smallskip}
  \hline
  \end{tabular}
  \begin{list}{}{}
  \item[$\A$] Pixel binning $2 \times 2$. \\
  \end{list}
  \end{flushleft}
  \end{table}


\subsection{Data reduction}

\noindent
The optical data were reduced using the MIDAS software package, developed by ESO.
Following the standard reduction procedures (bias subtraction, flat fielding with sky flats)
the remaining gradients in the background of galaxies that covered a major fraction of
the field of view were removed using a two-dimensional polynomial.
For some of the frames that were affected by bad columns these columns were also removed
by the standard MIDAS fitting routine. In order to increase the signal-to-noise ratio S/N
(important for an investigation of faint disk features) the images of fainter objects were
binned ($2 \times 2$).

Finally, the frames were rotated in such a way that the galaxy planes are in a horizontal
position (assuming symmetrical light distribution of the vertical disk profiles).
It should be stressed that the remaining small rotation error -- which is typically $\pm 0.4 \degr$
for most of the relatively uniform disks of non-interacting galaxies -- can be considered by
the disk fitting routine (Sect.~4). For galactic disks affected by strong disk warping
(mostly interacting objects) precise rotation was more difficult. For these cases, only the
inner parts of the galaxy disk were considered to determine the rotation.

For standard reduction of the near infrared data, the IRAF software package was used.
In particular, the sky frame subtraction, the flat fielding, and the combining of the
flat-fielded images was done with the ARNICA (Arcetri Near-Infrared Camera) add-on package.
The sky frames used for the flat field subtraction were obtained from a set of the nearest frames
in time, filtered by a median. The median filtering also removed the stars in the sky frames.
To produce a final source frame, the reduced, flat-fielded images were combined using
the ARNICA standard ``mosaic'' task (this task includes both median filtering, and centering
of frames by stars in the field).

Due to the small detector size -- the resulting field of view was $\approx 3'$ on average
(Table~\ref{telescopes}) -- images of larger fields were produced by mosaicing. Since this
is a time-expensive procedure, it was only applied in order to obtain images of some larger
objects. For precise adjustment of each of the frames we used either stars in
the field or the sharp central bulge regions of the galaxies themselves.

Image rotation was applied as explained for the optical images.

Isophote maps of the complete samples of interacting/merging and non-interacting galaxies,
respectively, are shown in Figs.~3 and 4 of this paper.


\section{Disk modelling}

\subsection{Overview on the disk model properties}

\noindent
In order to analyze and to compare the radial and vertical disk structure of a large
sample of highly-inclined/edge-on spiral galaxies, it is necessary to apply a disk
model that enables both a good quantitative and very flexible description of the
3-dimensional luminosity distribution. Although mathematical simplicity is also a
desired property of disk models, there should be a firm physical basis.

Therefore, a disk modelling- and fitting procedure was developed based on the results
of a fundamental study of edge-on spiral galaxies by van der Kruit \& Searle
(\cite{kruit1981a},\cite{kruit1981b}; \cite{kruit1982a},\cite{kruit1982b}) as well as
on other observational studies thereafter (e.g. van der Kruit \cite{kruit1988};
Barteldrees \& Dettmar \cite{barteldrees1994}; de Grijs \& van der Kruit \cite{grijs1996};
Just et al. \cite{just1996}). The disk model presented here also considers the effects
of an inclined disk ($i\neq 90 \degr$) as well as 3 different vertical luminosity
distributions.

\noindent
Using the following notation for disk model parameters

\begin{tabbing}

$L_{0}$   \hspace{15mm} \= $\ldots \,$ \= central luminosity density     \\
$i$                     \> $\ldots \,$ \> inclination angle of the disk  \\
$\Rmax$                 \> $\ldots \,$ \> disk cut-off Radius \\
$h$                     \> $\ldots \,$ \> disk scale length \\
$\zo$                   \> $\ldots \,$ \> disk scale height \\
$f_{n}(z,\zo)$          \> $\ldots \,$ \> vertical distribution, see (2) to (4) \\
$\Theta(\Rmax-r)$       \> $\ldots \,$ \> truncation (Heaviside-) function \\
                        \>             \> (1 for $ r < \Rmax $; 0 for $ r \geq \Rmax $)
\end{tabbing}

\noindent
the 3-dimensional luminosity distribution $L_{n}$ ($n=1,2,3$) of the disk can be
described in cylindrical coordinates by

\begin{equation}
L_{n}(r,z) \; = \; L_{0} \; \exp (-r/h) \; \; f_{n}(z,\zo) \;\; \Theta(\Rmax-r) \; .
\end{equation}

\noindent
A set of 3 functions $f_{n}(z,\zo)$ with the same asymptotic behaviour for $ z/\zo \gg 1 $
-- proposed by van der Kruit (\cite{kruit1988}); Wainscoat et al. (\cite{wainscoat1989},
\cite{wainscoat1990}); Burkert \& Yoshii (\cite{burkert1996}) -- is used to describe the
vertical luminosity distribution $L(z)$:

\begin{equation}
f_{1}(z,\zo) \; = \; 4 \, \exp  \; (-2 \, |z|/\zo) \; , 
\end{equation}

\begin{equation}
f_{2}(z,\zo) \; = \; 2 \; \sech \; (2 \, z/\zo) \; , 
\end{equation}

\begin{equation}
f_{3}(z,\zo) \; =  \;\;\;\, \iso  \; (z/\zo) \; . 
\end{equation}

\noindent
Thus, for large $z$, a comparison between different scale heights is possible via

\begin{equation}
\zo_{\, (\rm sech^{2})} \; = \; \sqrt{2} \; \zo_{\, (\rm sech)} \; = \; 2 \; \zo_{\, (\rm exp)} \; .
\end{equation}

\noindent
The use of this combination of 3 different vertical luminosity distributions meets the
mentioned specifications and allows also very flexible description of a large variety
of vertical profiles of galactic disks.

In order to obtain a final disk model, i.e. a two-dimensional intensity distribution,
integration of (1) along the line-of-sight through the disk is required. But there
exist only two possible disk models -- the projection face-on ($i=0 \degr$) and edge-on
($i=90 \degr$), combined with the isothermal disk model (4) -- which would allow an
analytical solution of (1) using a modified Bessel function (van der Kruit \& Searle
\cite{kruit1981a}). Therefore, integration of (1) must be calculated numerically.
%
%
Hence, the intensity $I(Y,Z)$ of one point (pixel) of the disk -- as it can be seen by an
observer at the projected coordinates $Y$ and $Z$ on the CCD -- can be written in cartesian
coordinates in the following form (for example, only the double exponential disk model (2)
is given here in its explicit form, for $i \not= 0 $; integration of models (3) and (4) is
analogous):

\begin{eqnarray}
I(Y,Z) \, = \, 4 \: L_{0} \: \int \limits^{r_{\rm max}}_{r_{\rm min}} \: \:
\exp \Bigg(- \: \frac{\sqrt{r^{2} \sin^{2}(i) + Y^{2}}}{h} \, \Bigg) \: \: \nonumber\\
\hspace{5mm} * \exp \Bigg(- \: \frac{2 \; (|r \cos(i) + Z / \sin(i)|)}{z_{0}} \, \Bigg) \; dr \; .
\end{eqnarray}

\noindent
The integration limits are, in a direction of the line-of-sight, given by the geometry of
the ``galaxy-cylinder'':

\begin{equation}
r_{\rm max,min} \; = \; \pm \; \sqrt{\, \Rmax^{2} - Y^{2}} \; / \; \sin(i) \;\; ,
\end{equation}

\noindent
and can be approximated for $i \approx 0 \degr$ by

\begin{equation}
r_{\rm max,min} \; = \; \pm \; 10 \: \zo \, ,
\end{equation}

\noindent
i.e. by a cylinder with a height of about 10 scale heights. At larger distances above
the disk plane the contribution of luminosity, acc. to (2)--(4), becomes negligibly.
The integration of (6) along the line-of-sight through the disk-``cylinder'' is
realized by a modified Newton-Cotes rule.
%

\noindent
According to $(1), (6)$, and (7) the intensity $I_{n}$ of any disk model can be
described by an integral of a family of 3 different density laws $L_{n}$, depending
on 5 free parameters each:

\begin{equation}
I_{n}(Y,Z) \; = \; \int \; L_{n} \, (L_{0},i,\Rmax,h,\zo) \; \, dr \; .
\end{equation}

\noindent
In practice, this is still a large parameter space. It therefore requires a further,
step-by-step restriction, realized in the following disk fitting procedure.




\begin{figure*}[t]

\vspace{119mm}

\hspace*{15mm}
\begin{minipage}[b]{14cm}
\begin{picture}(14,10)
{\includegraphics[angle=90,viewport=17 20 585 730,clip,width=150mm]{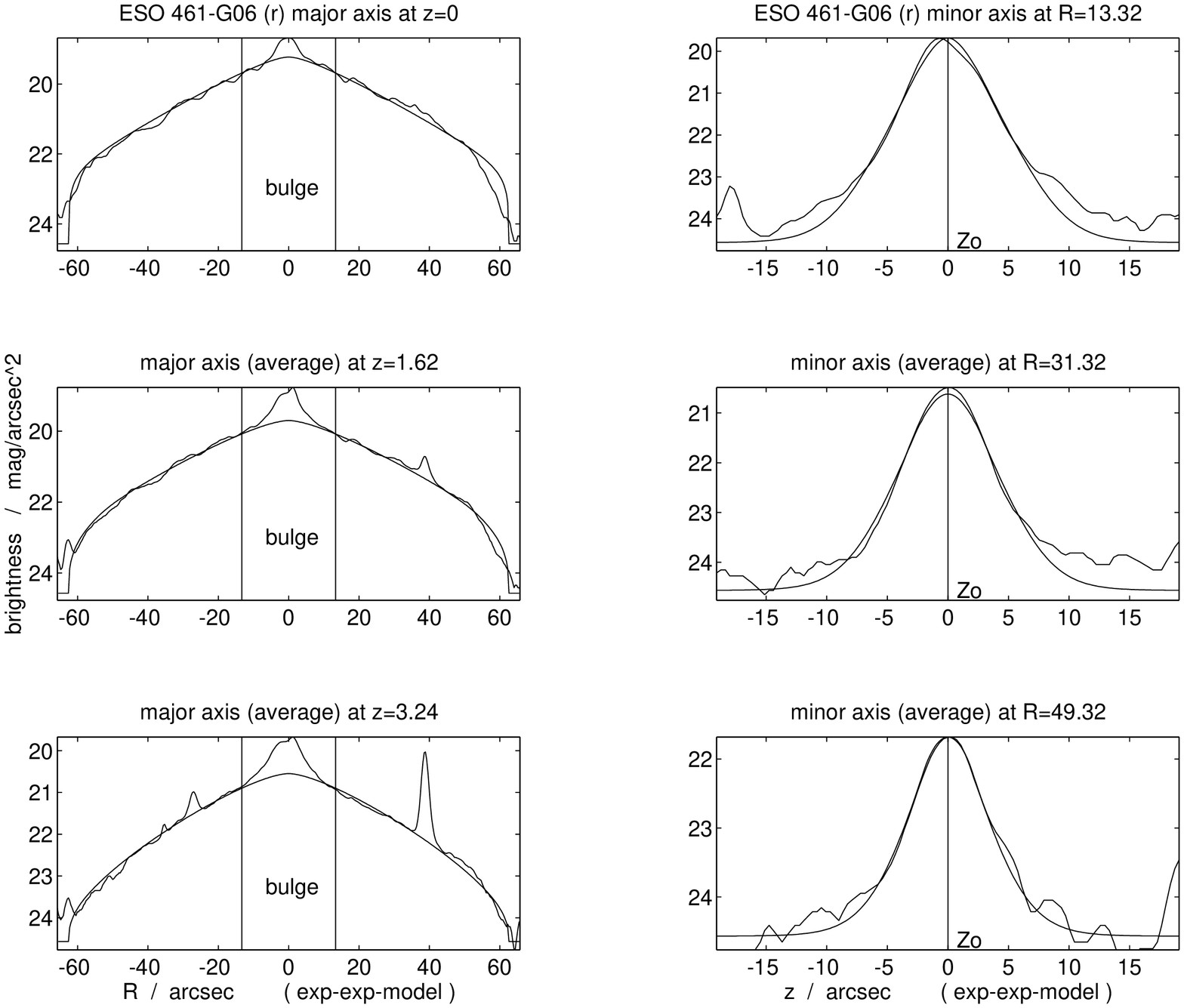}}
\end{picture}
\end{minipage}
\\
{\bf \noindent Fig.~2.} Example of the disk fitting procedure: Set of typical radial
(left panels) and vertical (right panels) profiles used for disk fitting of the Sc-galaxy
ESO 461-G06 ($r$-band, see also Fig.~5). The galaxy profiles (averaged) are displayed
together with the
final disk model, the used parameters are: $i=88.5\degr; \Rmax=60\farcs8; h=16\farcs0;
\zo=3\farcs0; f(z) \propto \rm exp$ (further explanation see text).

\end{figure*}



\begin{figure*}[t]

\vspace{45mm}

\begin{minipage}[b]{6cm}
\begin{picture}(8,7.3)
{\includegraphics[angle=0,viewport=020 20 590 600,clip,width=62mm]{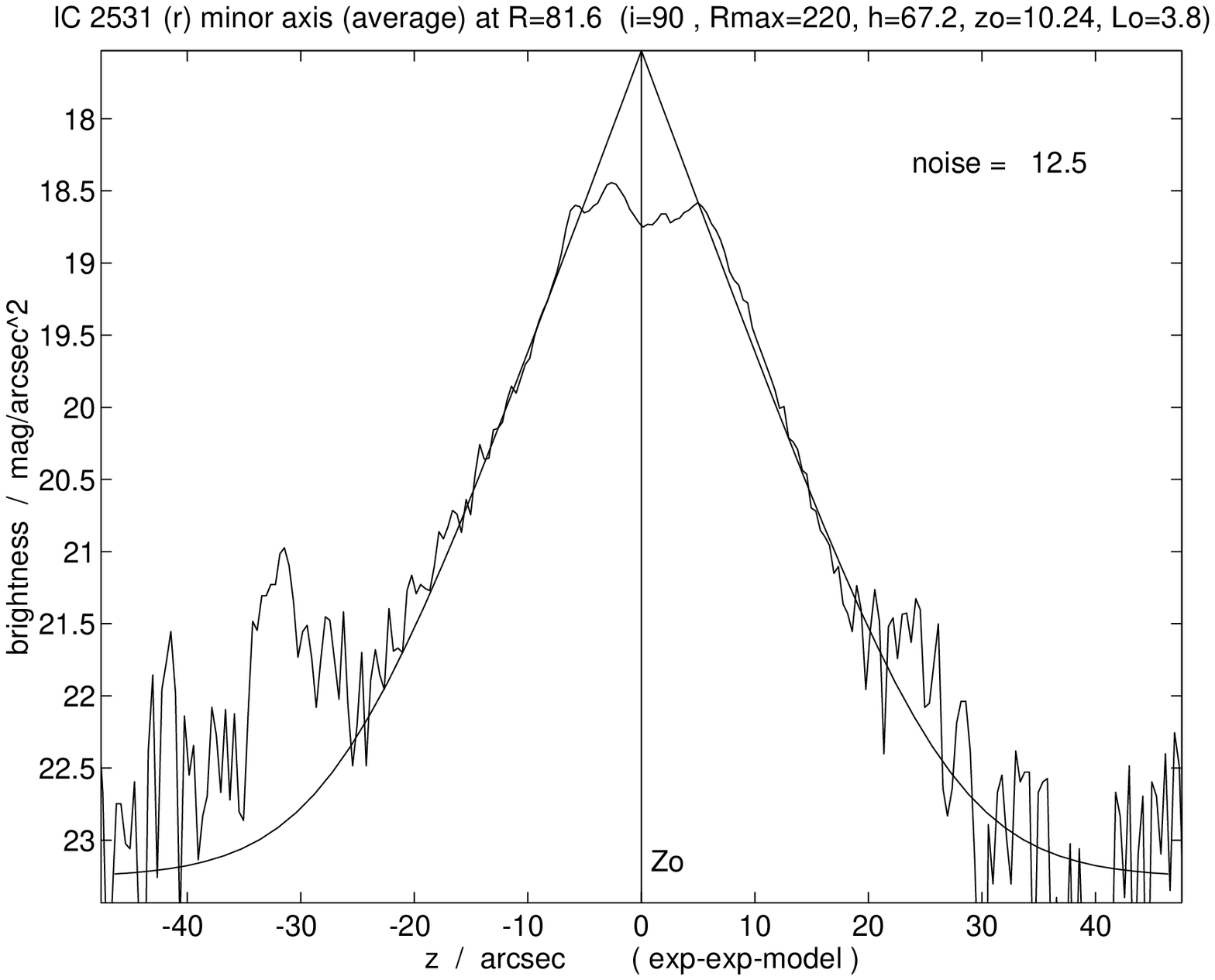}}
\end{picture}
\end{minipage}
\begin{minipage}[b]{6cm}
\begin{picture}(8,7.3)
{\includegraphics[angle=0,viewport=020 20 590 600,clip,width=62mm]{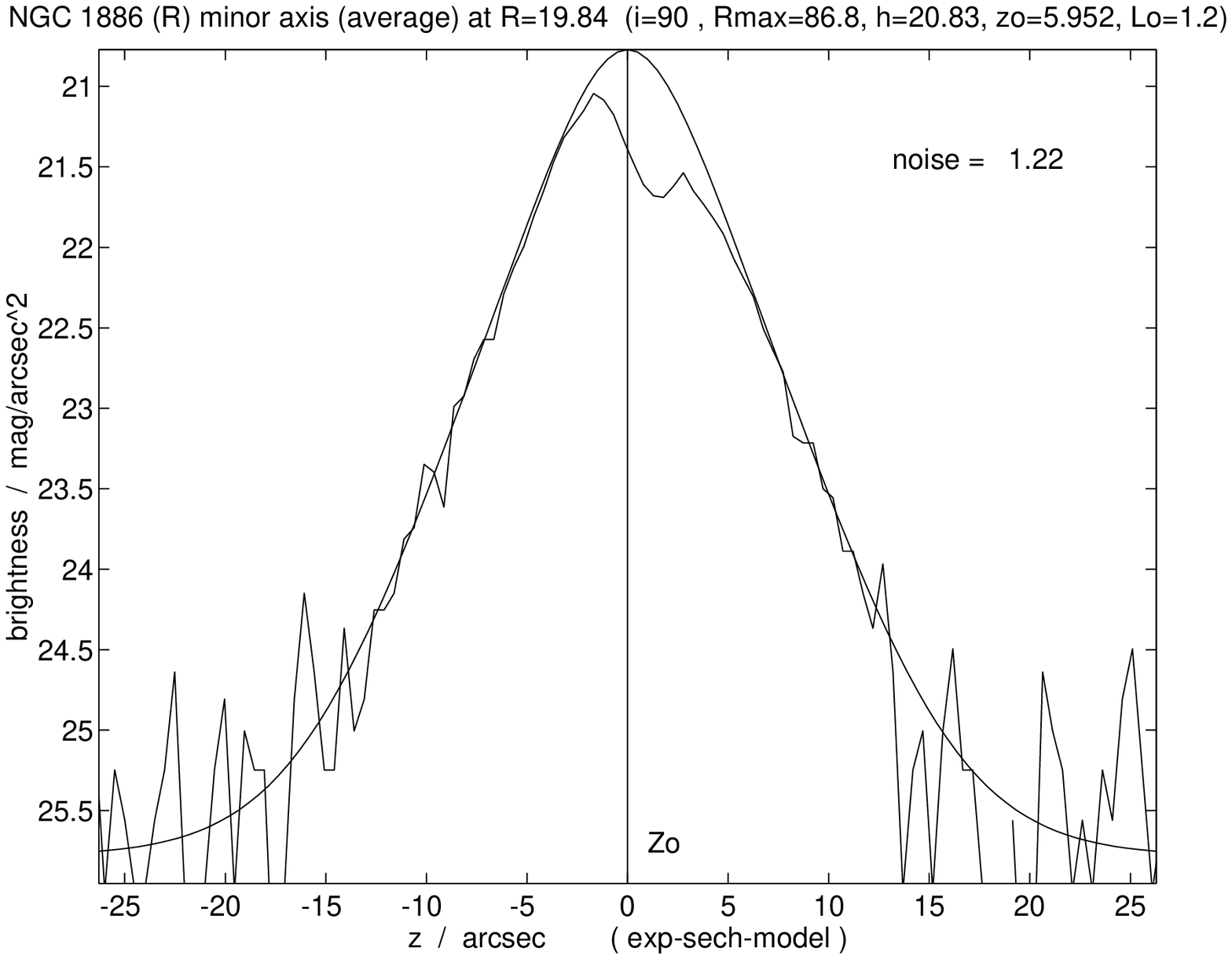}}
\end{picture}
\end{minipage}
\begin{minipage}[b]{6cm}
\begin{picture}(8,7.3)
{\includegraphics[angle=0,viewport=020 20 590 600,clip,width=62mm]{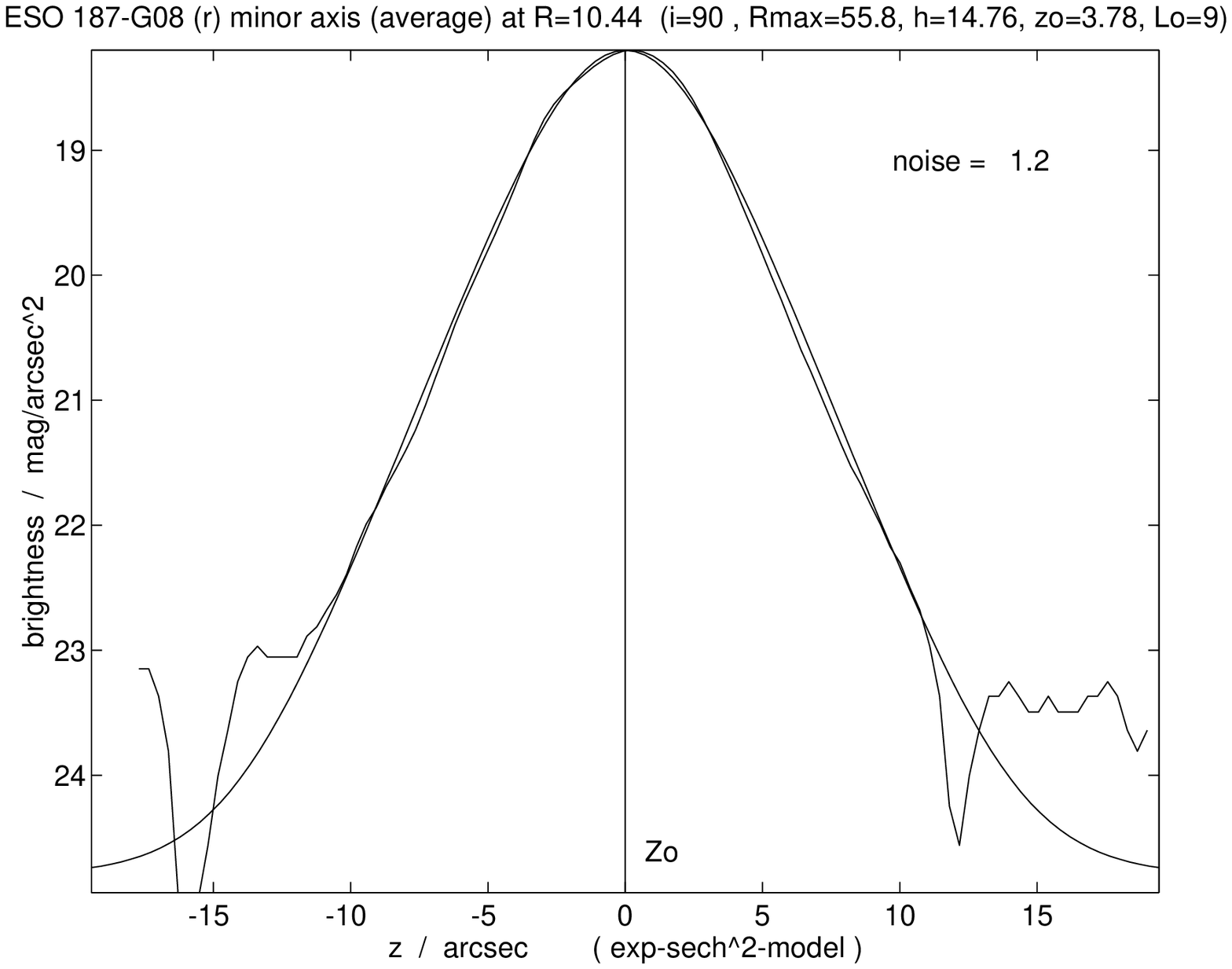}}
\end{picture}
\end{minipage}

\vspace{46mm}

\begin{minipage}[b]{6cm}
\begin{picture}(8,7.3)
{\includegraphics[angle=0,viewport=020 20 590 600,clip,width=62mm]{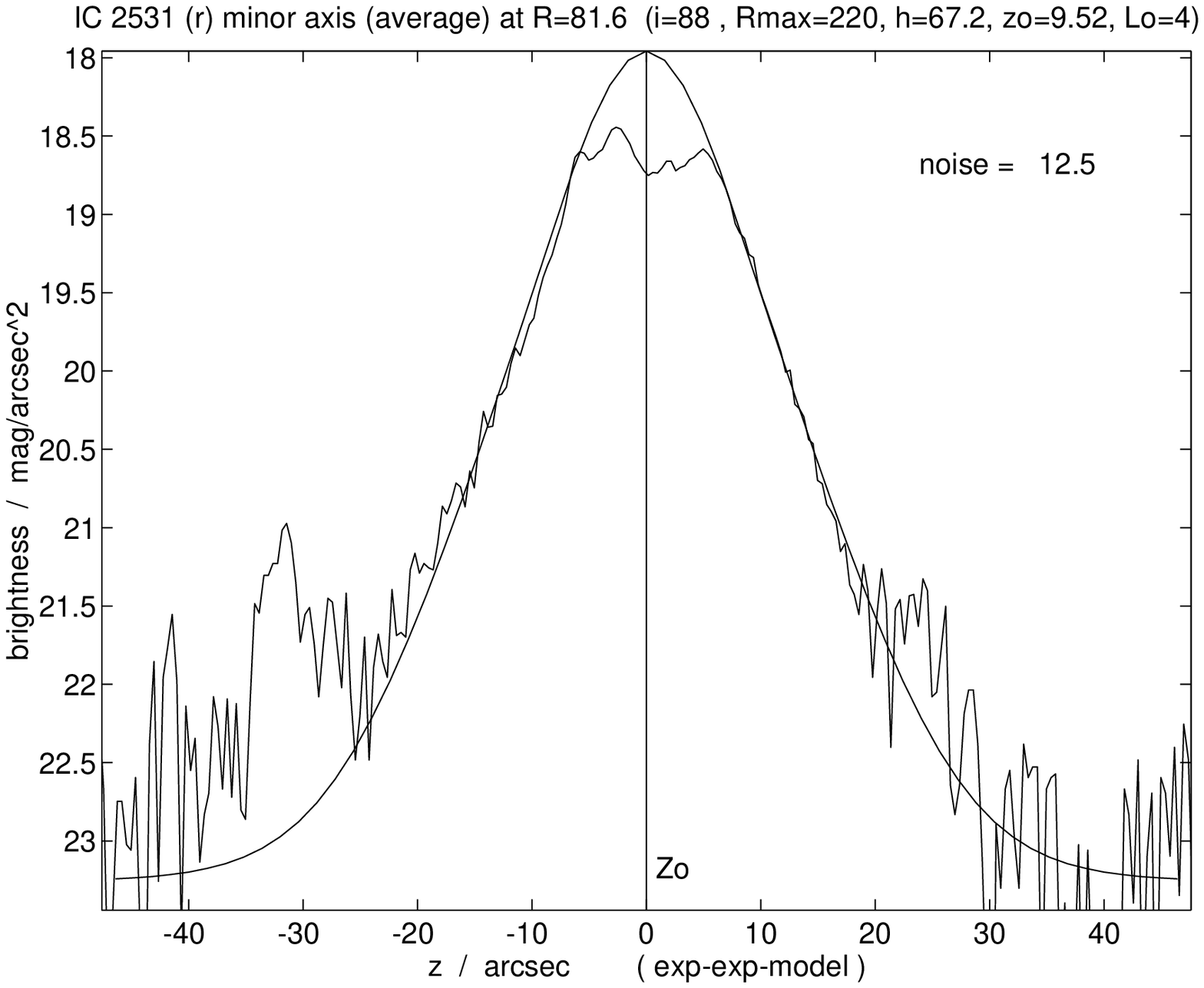}}
\end{picture}
\end{minipage}
\begin{minipage}[b]{6cm}
\begin{picture}(8,7.3)
{\includegraphics[angle=0,viewport=020 20 590 600,clip,width=62mm]{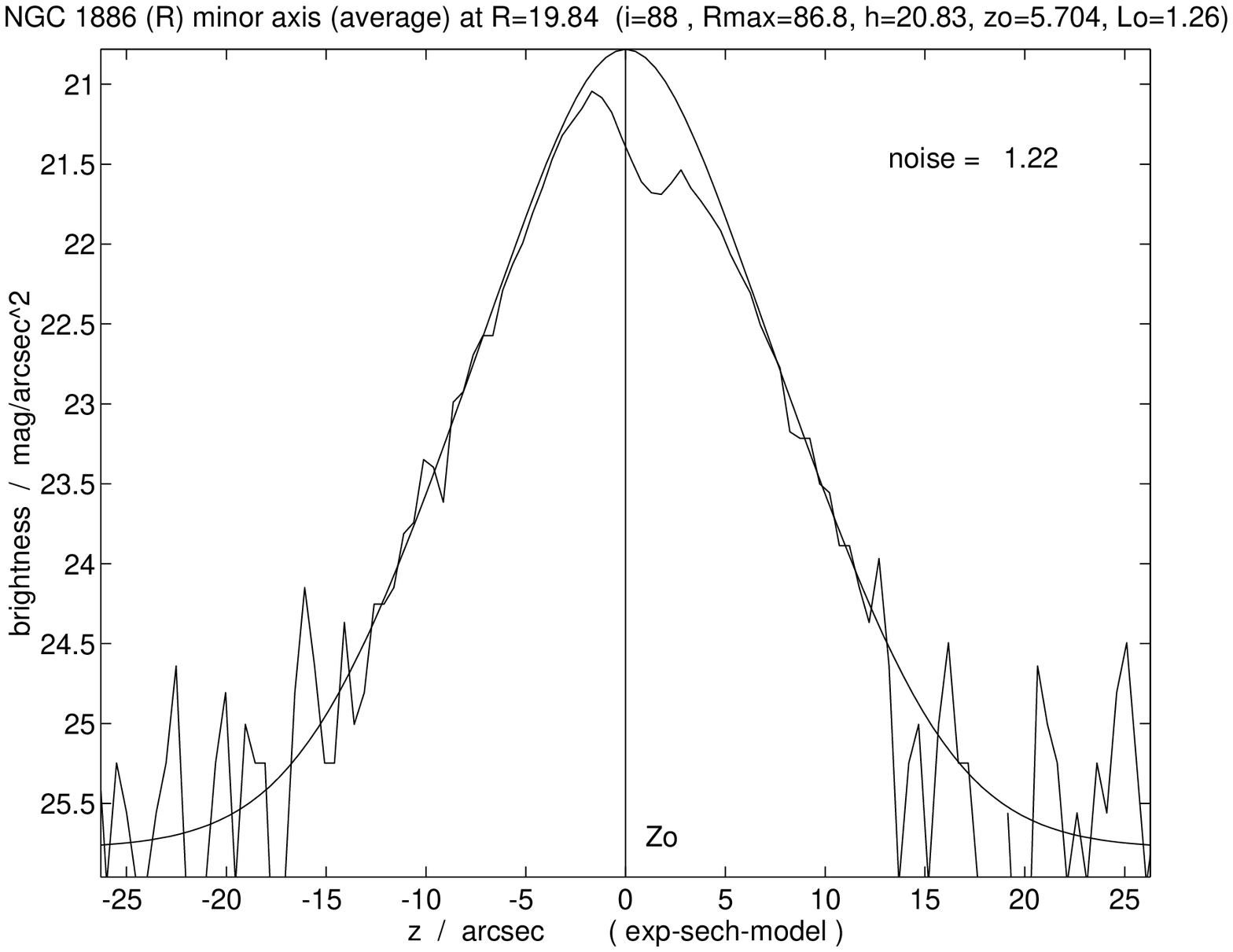}}
\end{picture}
\end{minipage}
\begin{minipage}[b]{6cm}
\begin{picture}(8,7.3)
{\includegraphics[angle=0,viewport=020 20 590 600,clip,width=62mm]{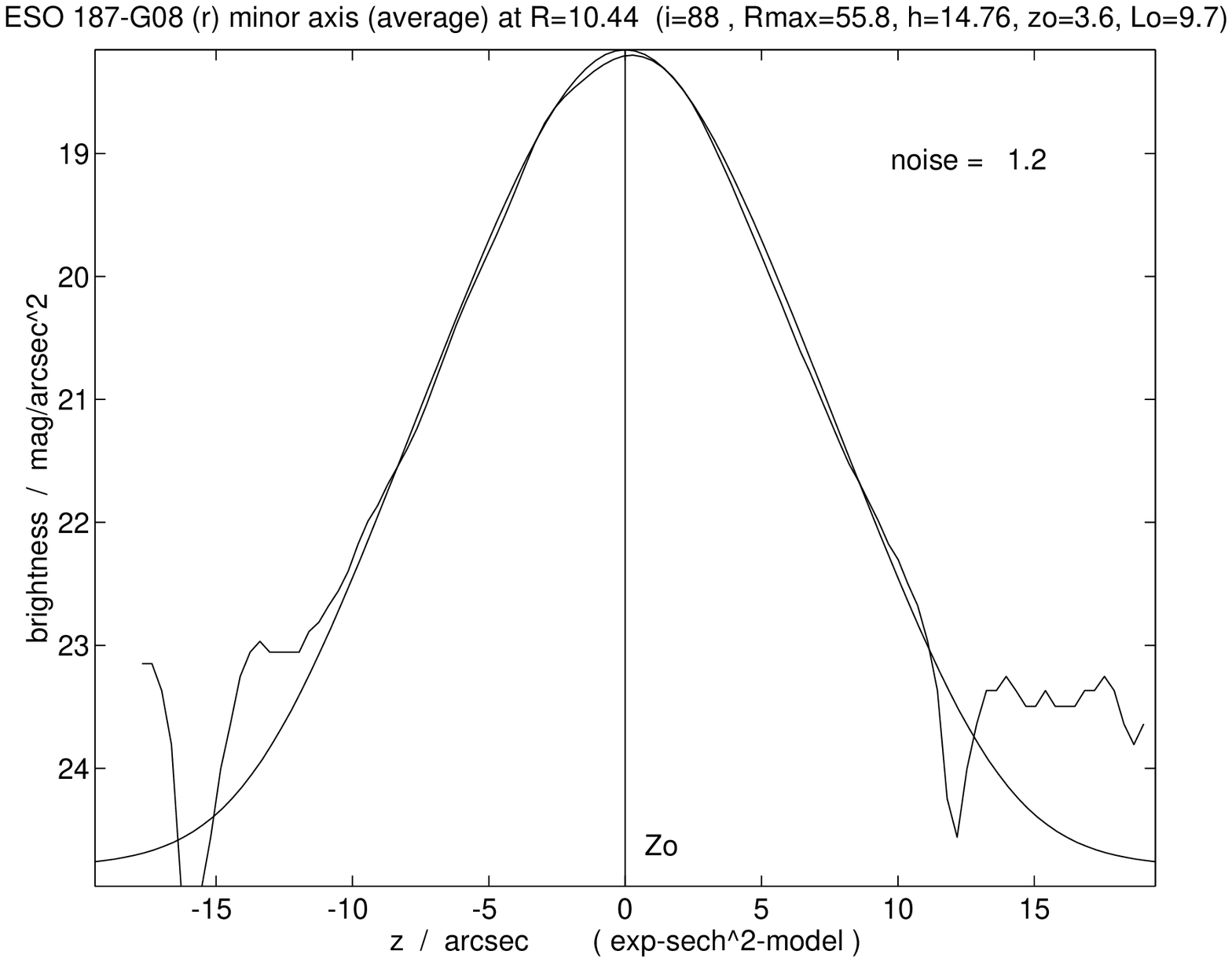}}
\end{picture}
\end{minipage}

\vspace{46mm}

\begin{minipage}[b]{6cm}
\begin{picture}(8,7.3)
{\includegraphics[angle=0,viewport=020 20 590 600,clip,width=62mm]{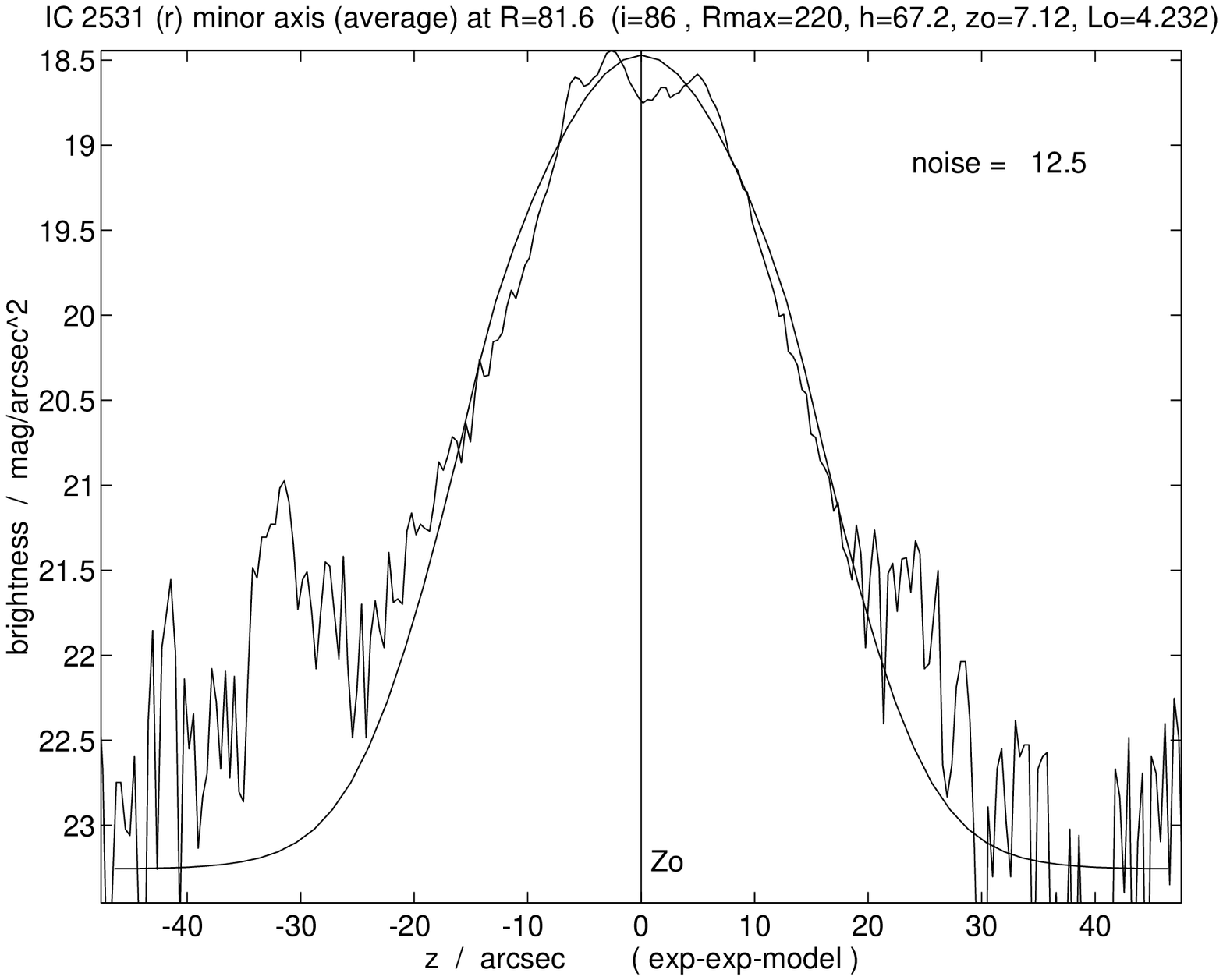}}
\end{picture}
\end{minipage}
\begin{minipage}[b]{6cm}
\begin{picture}(8,7.3)
{\includegraphics[angle=0,viewport=020 20 590 600,clip,width=62mm]{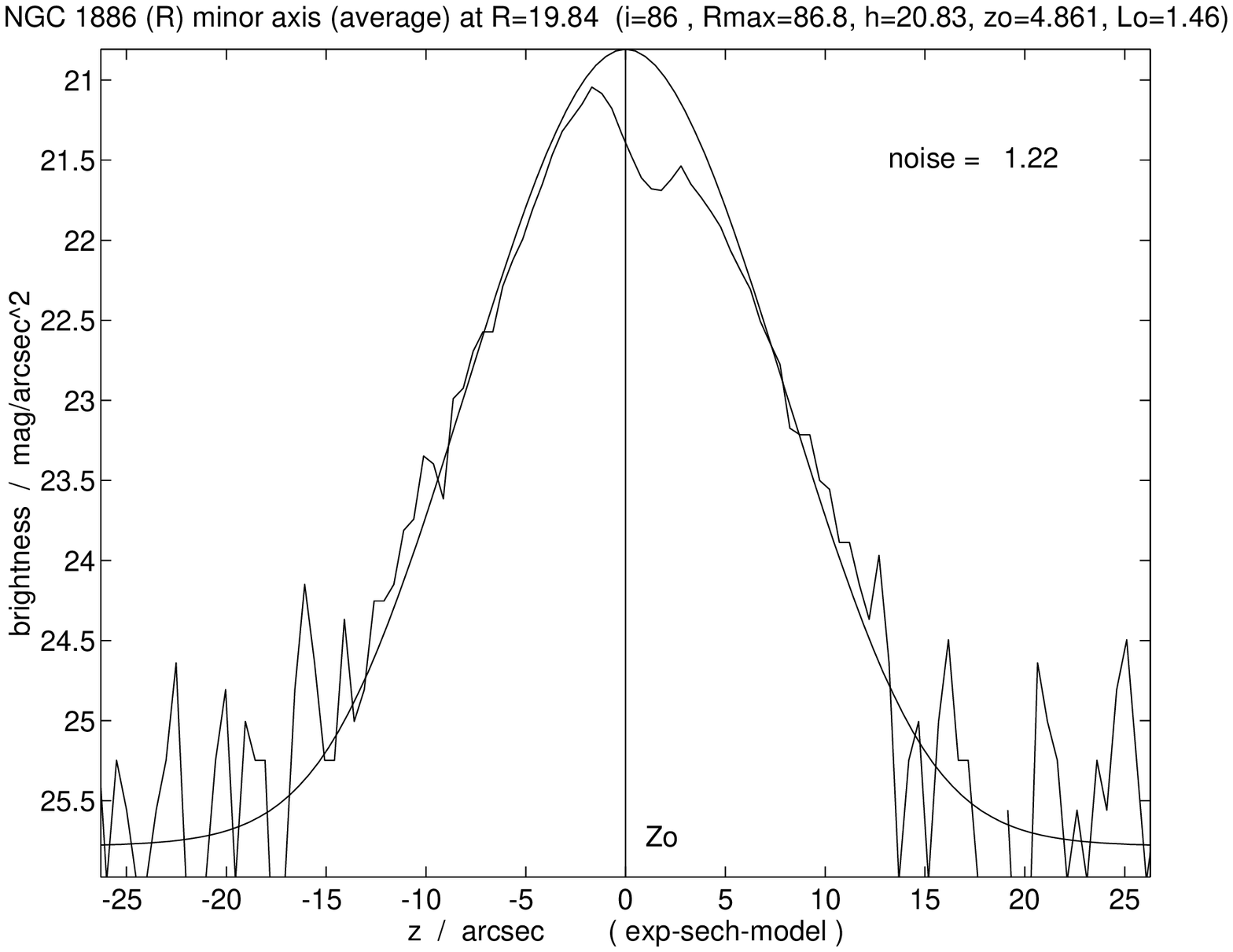}}
\end{picture}
\end{minipage}
\begin{minipage}[b]{6cm}
\begin{picture}(8,7.3)
{\includegraphics[angle=0,viewport=020 20 590 600,clip,width=62mm]{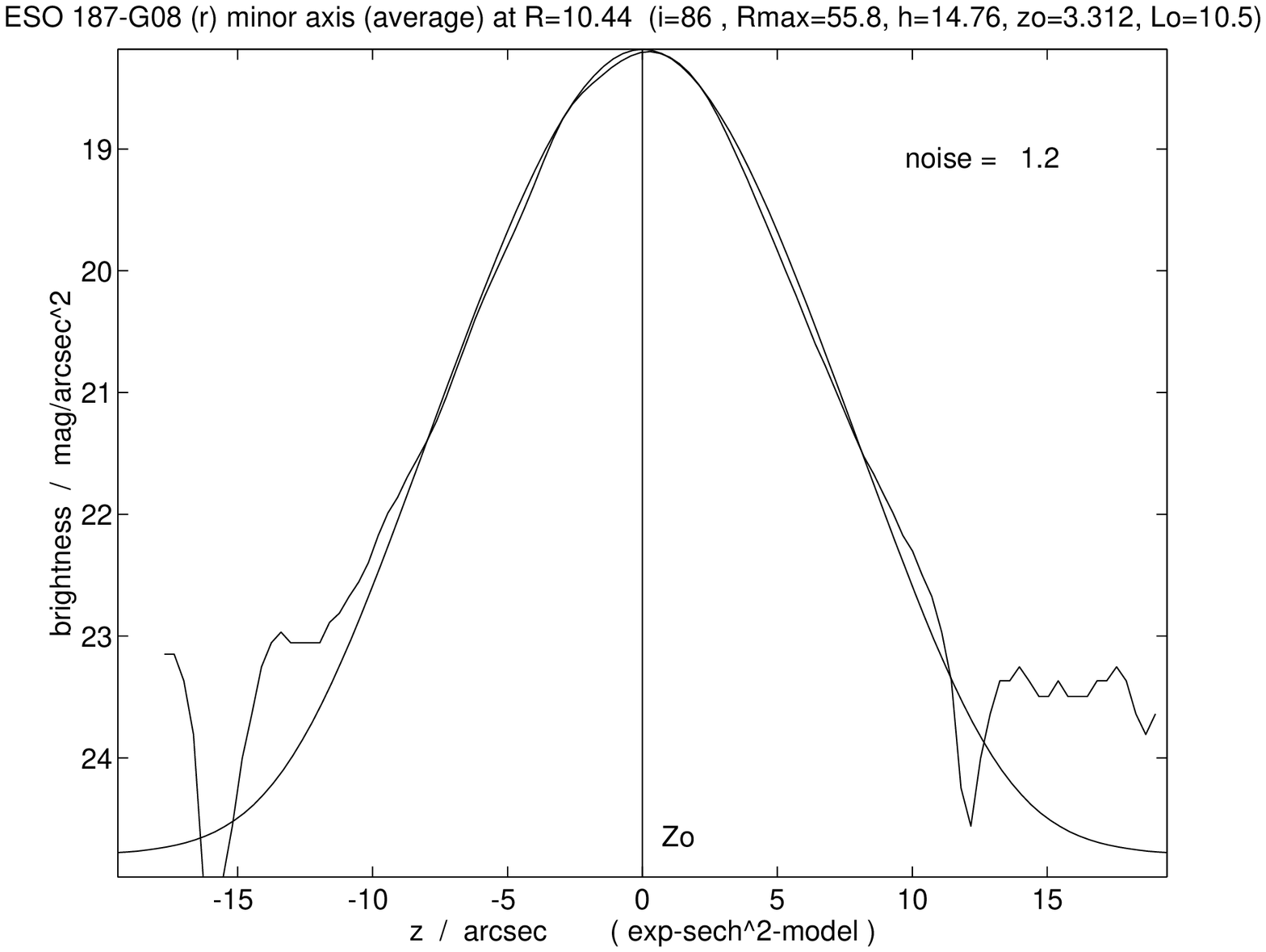}}
\end{picture}
\end{minipage}

\vspace{5mm}

{\bf \noindent Fig.~3.} The influence of inclination on the shape of vertical disk
profiles, shown for 3 different edge-on galaxies and their corresponding models
(columns from left to right): $f(z) \propto$ exp (IC 2531); $f(z) \propto$ sech
(NGC 1886); $f(z) \propto \iso$ (ESO 187-G08). 
Although the disk inclination is varied between $i=86\degr ... \, 90\degr$, for each of
the vertical profiles a quantitative good fit can be found by changing only the scale
height $\zo$ and the central luminosity density $L_{0}$ (as indicated; further
explanation see text).
\end{figure*}


\subsection{The disk fitting procedure}

In order to derive relible disk parameters we developed two independent disk fitting procedures
that were designed to comply the special tasks proposed for the first (Paper~I+II) and second
(Paper~III) part of this study (see Sect.~1).

The first fitting procedure was realized semi-automatically. It uses different graphs enabling
a direct comparison by eye of a set of radial and vertical profiles within a preselected
disk area with those of an underlying disk model. To take advantage of symmetrical light
distribution of galactic disks the profiles used for modelling are averaged over two
quadrants. They are usually displayed equidistantly and cover the whole fitting area.
The size of the fitting area vastly depends on the properties of the individual image,
i.e. on both the image quality and some special features of the galaxy disk itself.
Therefore the program allows -- after a first inspection
of the profiles -- an interactive selection of a qualified fitting area. Depending on the S/N
ratio and the spatial resolution of the images it is also possible to use a desired pixel
binning for both image and model (as an example Fig.~2 shows one of the sets of radial and
vertical disk profiles). This kind of modelling allows fast and flexible, but reliable disk
fitting of a large number of galaxies -- in particular of interacting/merging galaxies -- 
whose disk profiles do often show considerable deviations from the simple model. Such galaxies
are therefore difficult to handle with conventional least squares fitting methods.

The second fitting procedure uses the same set of disk models as described before but a least
square fitting in order to fit the scale height as a function of the galactocentric distance.
The properties of the program will be described in detail in Paper~III.
To ensure homogenity of the fitting results the fit quality reached with this method was
compared with that of the above described semi-automated disk modelling (for this purpose
the data of non-disturbed disk profiles of galaxies in the non-interacting sample were used).
It was found that both methods give consistent results with errors on (or below) a 10\%-level.
The errors found for the fits of the scale height only are even smaller than 5\%.
The remaining discrepancies are not due to the individual method itself but rather to the
following two reasons: first, the fitting areas are slightly different (i.e. they are usually
more restricted for the least square method as a result of its sensibility against the error
sources described in the following). Second, the scale height of a large fraction of galaxy
disks investigated shows variations of different absolute size (both irregularly and systematically,
i.e. gradients) along the galactocentric distance. Therefore this point will be discussed
in detail in Paper~III. A detailed comparison of the semi-automated disk modelling with
another, independent developed least square fitting routine will be given in Pohlen et al.
(\cite{pohlen2000}).

After comparison of both fitting methods we choose to use the semi-automated disk modelling
for this part of the study because it combines the advantages of fast and flexible fitting
with a high accuracy. In view of the existing data the point of flexible fitting is very
important since -- despite all structural similarities -- the radial and vertical profiles
of induvidual (disk) galaxies are unique. The profiles investigated here are often heavily
contaminated  by light from, e.g., a bulge and/or a bar, a nearby companion, other disk
components, foreground stars or reflections from bright stars.
In addition, the modelling may be complicated by strong dust extinction along the galactic
plane, by a low S/N ratio, or a considerably warped disk (mainly interacting/merging
galaxies). Since most of the foregoing deviations can not be easily quantified their complete
consideration by an automated fitting procedure thus seems almost impossible and would cause
unpredictably errors.

In the following step-by-step procedure the disk parameters of the semi-automated disk
fitting procedure are reduced systematically:
as a first step, the inclination is determined by using the axis ratio of the dust lane in
the disk plane of optical images. Given a relatively sharp dust lane, an accuracy of less
than $\pm 0.5 \degr$ can be reached. The central luminosity density, $L_0$, is calculated
automatically by the fitting program for each new parameter set by using a number of
preselected reference points along the disk (outside the bulge- or bar-light contaminated
regions).

Given a sufficient S/N ratio, the cut-off radius $\Rmax$ can be fitted to the major axes
profiles with an accuracy between $(5-10)\%$. The cut-off is determined by a significant
decrease of the intensity extrapolated to $I=0$ (left panels in Fig.~2). For highly-inclined
disks such as the ones studied here, the effect of a variation of $\Rmax$ on the slope of
radial disk profiles is negligible.

Thus, the remaining ``real'' fitting parameters are the disk scale length and height, $h$ and
$\zo$, as well as the set of 3 functions $f_{n}(z)$. Within the following procedure both the
scale length and the scale height are fitted in an iterative process until a first good
convergence of the ``global'' fit is achieved. During this process, the quality of the
corresponding fits can be checked simultaneously using a small set of radial and vertical
disk profiles (usually 3 profiles each, see Fig.~2 as an example). For further small
corrections, if necessary, both parameters can be considered as independent and thus
separated without any loss of accuracy. For this ``fine tuning'' a set with more disk
profiles (usually 6-8) is used for both parameters.

During the iterative process of modelling the scale length is fitted to a set of averaged
major axes profiles in a radial region typically between $(0.7 - 2.8)h$ and vertically
outside strong dust extinction (left panels in Fig.~2). The fit quality of the vertical
profiles along the disk can be used as a cross-check for the scale length (right panels
in Fig.~2). As a result of the error sources mentioned at the beginning the disk scale
length of a galaxy can be reproduced (with the same method) with an accuracy of $(5-10)\%$.
In contrast to this, the scale lengths derived with different methods can in some cases
differ by $(25-50)\%$.

The disk scale height $\zo$ is estimated by fitting the $z$-profiles inside the previously
selected radial regions, which are outside the bulge- or bar contamination (right panels in
Fig.~2). The vertical region used for the fits is typically between $(0.2 - 2)\zo$, but
depends strongly on the individual characteristic of the dust lane. If the disk inclination
is known precisely, this fitting method works reliably. Otherwise, additional errors may be
introduced (see next Sect.).

The vertical disk profiles of most of the galaxies investigated in optical passbands enable
a reliable choice of the quantitatively best fitting function $f_{n}(z)$. This is because
deviations between different models become visible at vertical distances larger than that
of the most sharply-peaked dust regions. For those disks that are affected by strong dust
extinction, the choice was made easier by using a combination of both optical and near
infrared profiles. For these cases profiles of both passbands were, if available, used
for fitting.


\subsection{The influence of inclination on vertical disk profiles}

If a large sample of disk galaxies is investigated statistically -- as is the case in this
study -- moderate deviations from edge-on orientation of the order of $\pm 5\degr$ are common
(Sect.~2). It was found in this study and in Schwarzkopf \& Dettmar (\cite{schwarzkopf1998})
that, if reliable values for the scale height are desired, deviations from $90\degr$ larger
than $4\degr - 5\degr$ can not be neglegted. Therefore in this section the influence of small
changes of inclination on vertical disk profiles and on the resulting parameters will be
investigated in detail.

For that purpose we selected 3 $r$-band images of galaxies in the non-interacting sample with
well known disk parameters, but with different vertical profiles: $f(z) \propto$ exp (IC 2531);
$f(z) \propto$ sech (NGC 1886); and $f(z) \propto \iso$ (ESO 187-G08). At 3 different
inclinations ($i=90\degr$; $88\degr$; and $86\degr$) the vertical profiles of each model
were best-fitted to the observed disk profiles (see left, middle, and right panels of Fig.~3).
To compensate for the effect of inclination, both the disk scale height, $\zo$, and the
central luminosity density, $L_{0}$, must be changed. As can be seen in the disk parameters
(middle raw in Fig.~3 and Table~\ref{inclination}) the effect for $\zo$ is, at $i=88\degr$,
around the $5\%$-level for all vertical models, whereas for $i=86\degr$ (Fig.~3 bottom) the
error amounts to $\approx 30\%$ for the exp-model and, after all, to $\approx 13\% - 18\%$ for
the $\iso$ and sech-models. At smaller inclinations the quality of all vertical fits decreases
rapidly and allows no more qualitative good fits.
As expected, the effect of slight changes of inclination on both the scale height and the
shape of vertical profiles near the disk plane is the strongest for the exp-model
(left panels in Fig.~3).

The effect shown above is due to the fact that the slope of the vertical disk profiles -- in
a region that is relevant for fitting (previous section) -- remains nearly unchanged in the
range between $i \approx 85\degr - 90\degr$, whereas this is not true for the width of
the vertical profiles. Hence, to obtain reliable disk parameters for $\zo$ and $L_{0}$ it is
required that both

\begin{itemize}
\item the inclination of the disk is known precisely (within $\pm2\degr$) and

\smallskip

\item a disk model with a flexible inclination is used.
\end{itemize}

\noindent
Otherwise, substantial errors for $\zo$ and $L_{0}$ are introduced (Table~\ref{inclination}
lists averaged errors $\Delta \zo$ and $\Delta L_{0}$, obtained by fitting 5 galaxies with
3 different vertical disk models each).

%

%
%
%

\tabcolsep1.2mm

  \begin{table}[t]
  \caption[ ]
  {The influence of inclination on vertical disk profiles. \\
   columns: (1) Disk inclination used for the model; (2), (4), (6) Error of disk scale height
   using $f(z) \propto$ exp, sech, and $\iso$; (3), (5), (7) Error of central luminosity density
   using $f(z) \propto$ exp, sech, and $\iso$.}
  \label{inclination}
  \medskip
  \begin{flushleft}
  \begin{tabular}{ccrrccrrccrr}
  \cline{1-12}
  \hline\hline\noalign{\smallskip}
   \multicolumn{1}{c}{Disk}                     &&  \multicolumn{2}{c}{$f(z) \propto \rm exp$} &&&
   \multicolumn{2}{c}{$f(z) \propto \rm sech$}  &&& \multicolumn{2}{c}{$f(z) \propto \iso$}    \\
   \multicolumn{1}{c}{inclination}              &&  \multicolumn{1}{c}{$\Delta \zo$}   &
   \multicolumn{1}{c}{$\Delta L_{0}$}           &&& \multicolumn{1}{c}{$\Delta \zo$}   &
   \multicolumn{1}{c}{$\Delta L_{0}$}           &&& \multicolumn{1}{c}{$\Delta \zo$}   &
   \multicolumn{1}{r}{$\Delta L_{0}$}  \\
   \noalign{\smallskip}
   \multicolumn{1}{c}{(1)} && \multicolumn{1}{c}{(2)} &   \multicolumn{1}{c}{(3)} &&&
   \multicolumn{1}{c}{(4)} &  \multicolumn{1}{c}{(5)} &&& \multicolumn{1}{c}{(6)} &
   \multicolumn{1}{r}{(7)} \\
  \hline\noalign{\smallskip}
  $88\degr$   &&   $ 7\%$   &   $ 5\%$   &&&   $ 5\%$   &   $ 5\%$   &&&   $ 5\%$   &   $\;\;7\%$  \\
  $86\degr$   &&   $30\%$   &   $11\%$   &&&   $18\%$   &   $18\%$   &&&   $13\%$   &   $   18\%$  \\
  \noalign{\smallskip}
  \hline
  \end{tabular}
  \end{flushleft}
  \end{table}


\section{Summary and conclusions}

\noindent
Optical and near infrared photometric data of a sample of 110 highly-inclined/edge-on
disk galaxies are presented. This sample consists of two subsamples of 61
non-interacting galaxies and 49 minor merging candidates. Additionally, 41
of these galaxies were observed in the near infrared.

The sample selection, observations, and data reduction are described.
We show that -- although the subsamples are naturally slightly polluted due
to unavoidable selection effects -- the distribution of their morphological types
is almost indistinguishable, covering the range between \linebreak
$0 \le T \le 9$.
This is important for the forthcoming detailed statistical study focused on the
influence of interaction and minor merger on the radial and vertical disk structure
of spiral galaxies.

Moreover, a 3-dimensional disk modelling- and fitting procedure is described in order to
analyze and to compare the disk structure of our sample galaxies by using characteristical
parameters. We find that the vertical brightness profiles of modelled galaxy disks respond
very sensitive even to small changes of inclination around perfect edge-on orientation.
Therefore, projection effects of highly-inclined disks must be considered.


\begin{acknowledgements}

This work was supported by {\em Deutsche Forschungsgemeinschaft}, DFG,
under grant no. GRK 118/2.
\\
This research has made use of the NASA/IPAC Extragalactic Database (NED).

\end{acknowledgements}




\newpage

%
%
%

\tabcolsep2.35mm

\begin{center}
  \begin{table*}
  \caption[ ]{Complete list of the optical/near infrared galaxy samples used in this study. \\
   columns: (1) Serial number; (2) Galaxy name; (3), (4) Center coordinates for epoch 2000,
   based on NASA/IPAC Extragalactic Database (NED); (5) Telescopes used: 2.2 CA=2.2m Calar Alto,
   2.2 ESO=2.2m ESO/MPIA, 1.5 DA=1.54m Danish, 1.2 CA=1.23m Calar Alto, 1.1 LO=1.1m Lowell,
   1.0 HL=1.06m Hoher List, 0.6 BO=0.6m Bochum; (6) Observing dates (mm dd yy); (7) Passband used;
   (8) Exposure time (total); (9) Total blue magnitude (if available); (10) Revised Hubble type,
   based on NED, according to Table~\ref{types}; (11) Notes: 1 = data of this paper,
   2 = supplementary data from Barteldrees \& Dettmar (\cite{barteldrees1994}).}
  \label{samples}
  \begin{flushleft}
  \begin{tabular}{rllllcccrrc}
  \cline{1-11}
  \hline\hline
  \noalign{\smallskip}
  \multicolumn{1}{r}{No.}          & \multicolumn{1}{c}{Galaxy}     &
  \multicolumn{1}{c}{RA (2000)}    & \multicolumn{1}{c}{Dec (2000)} &
  \multicolumn{1}{c}{Telescope}    & \multicolumn{1}{c}{Date}       &
  \multicolumn{1}{c}{Band}         & \multicolumn{1}{c}{Exp.-time}  &
  \multicolumn{1}{c}{$B_{\rm T}$}  & \multicolumn{1}{c}{Type}       & \multicolumn{1}{c}{Note}  \\
  \noalign{\smallskip}
   &  & \multicolumn{1}{l}{\hspace*{1mm}h\hspace{3mm}m\hspace{3mm}s}    &
   \multicolumn{1}{c}{\hspace*{1mm}$\degr$\hspace{3mm}$'$\hspace{3mm}$''$}  &  &
   &  & \multicolumn{1}{c}{[min]}  & \multicolumn{1}{c}{[mag]}   &  &  \\
  \noalign{\smallskip}
  \multicolumn{1}{c}{(1)}  &  \multicolumn{1}{c}{(2)}  &  \multicolumn{1}{c}{(3)} &
  \multicolumn{1}{c}{(4)}  &  \multicolumn{1}{c}{(5)}  &  \multicolumn{1}{c}{(6)} &
  \multicolumn{1}{c}{(7)}  &  \multicolumn{1}{c}{(8)}  &  \multicolumn{1}{c}{(9)} &
  \multicolumn{1}{c}{(10)} &  \multicolumn{1}{c}{(11)} \\
  \noalign{\smallskip}
  \hline
%
%
%
%
  \noalign{\medskip}
  \multicolumn{11}{c}{\bf Interacting / Merging} \\
  \noalign{\medskip}
  \hline
  \noalign{\medskip}
%
%
  1  & NGC 7       &  00 08 20.6  & $-$29 54 54  &  0.6 BO  & Dec 27 96  & $R$  &  75  & 14.40 &  4.5 & 1 \\
  2  & UGC 260     &  00 27 03.0  &   +11 35 03  &  1.0 HL  & Sep 04 97  & $R$  &  65  & 13.71 &  6.0 & 1 \\
  3  & NGC 128     &  00 29 15.0  &   +02 51 55  &  1.2 CA  & Sep 08 96  & $R$  &  30  & 12.77 & -2.0 & 1 \\
     &             &              &              &  2.2 CA  & Sep 03 96  & $K$  &   4  &  ---  & -2.0 & 1 \\
  4  & AM 0107-375 &  01 09 42.0  & $-$37 42 27  &  0.6 BO  & Dec 28 96  & $R$  &  65  &  ---  &  3.5 & 1 \\
  5  & ESO 296-G17 &  01 23 55.0  & $-$38 00 44  &  0.6 BO  & Dec 29 96  & $R$  &  60  & 16.38 &  3.0 & 1 \\
  6  & ESO 354-G05 &  01 52 07.0  & $-$33 31 46  &  0.6 BO  & Jan 05 97  & $R$  &  75  & 15.95 &  4.0 & 1 \\
  7  & ESO 245-G10 &  01 56 44.0  & $-$43 58 23  &  0.6 BO  & Dec 31 96  & $R$  &  75  & 14.28 &  3.0 & 1 \\
  8  & ESO 417-G08 &  02 58 47.0  & $-$32 05 52  &  0.6 BO  & Jan 04 97  & $R$  &  75  & 13.64 &  0.7 & 1 \\
  9  & ESO 199-G12 &  03 03 25.0  & $-$50 29 43  &  0.6 BO  & Dec 28 96  & $R$  &  88  & 15.52 &  8.0 & 1 \\
 10  & ESO 357-G16 &  03 19 34.0  & $-$32 27 53  &  0.6 BO  & Jan 02 97  & $R$  &  75  & 14.34 &  3.0 & 1 \\
%
%
 11  & ESO 357-G26 &  03 23 56.0  & $-$36 27 50  &  1.2 CA  & Jan 01 97  & $R$  &  75  & 11.37 & -1.0 & 1 \\
 12  & ESO 418-G15 &  03 39 23.0  & $-$31 19 19  &  0.6 BO  & Dec 28 96  & $R$  &  55  & 12.40 &  4.0 & 1 \\
 13  & NGC 1531/32 &  04 11 59.0  & $-$32 50 57  &  1.5 DA  & Apr 08 97  & $r$  &  30  & 10.65 &  2.7 & 1 \\
 14  & ESO 202-G04 &  04 17 46.0  & $-$50 09 51  &  0.6 BO  & Dec 26 96  & $R$  &  80  & 13.47 &  2.0 & 1 \\
 15  & ESO 362-G11 &  05 16 39.0  & $-$37 06 00  &  0.6 BO  & Jan 05 97  & $R$  &  75  & 13.04 &  4.0 & 1 \\
 16  & NGC 1888    &  05 22 35.0  & $-$11 29 58  &  1.5 DA  & Apr 08 97  & $r$  &  40  & 12.83 &  5.0 & 1 \\
     &             &              &              &  2.2 ESO & Apr 10 97  & $K'$ &  40  &  ---  &  5.0 & 1 \\
 17  & ESO 363-G07 &  05 33 13.0  & $-$36 23 59  &  0.6 BO  & Dec 30 96  & $R$  &  75  & 13.25 &  5.5 & 1 \\
 18  & ESO 487-G35 &  05 42 01.0  & $-$22 56 43  &  0.6 BO  & Jan 01 97  & $R$  &  90  & 13.39 &  7.8 & 1 \\
 19  & NGC 2188    &  06 10 10.0  & $-$34 06 22  &  1.5 DA  & Apr 09 97  & $r$  &  75  & 12.14 &  9.0 & 1 \\
     &             &              &              &  2.2 ESO & Apr 12 97  & $K'$ &  33  &  ---  &  9.0 & 1 \\
 20  & UGC 3697    &  07 11 21.3  &   +71 50 06  &  1.1 LO  & Mar 16 96  & $R$  &  33  & 13.50 &  7.0 & 1 \\
     &             &              &              &  1.2 CA  & Mar 01 96  & $H$  &  53  &  ---  &  7.0 & 1 \\
 21  & ESO 060-G24 &  09 02 40.3  & $-$68 13 38  &  1.5 DA  & Jun 03 98  & $r$  &  30  & 13.95 &  2.5 & 1 \\
%
%
 22  & ESO 497-G14 &  09 07 42.0  & $-$23 37 15  &  0.6 BO  & Dec 26 96  & $R$  &  75  & 14.16 &  3.0 & 1 \\
 23  & NGC 2820    &  09 21 47.0  &   +64 15 29  &  1.2 CA  & Mar 01 96  & $H$  &  53  & 13.28 &  5.0 & 1 \\
 24  & NGC 3044    &  09 53 40.0  &   +01 34 46  &  1.5 DA  & Apr 08 97  & $r$  &  45  & 12.46 &  5.0 & 1 \\
     &             &              &              &  2.2 ESO & Apr 10 97  & $K'$ &  40  &  ---  &  5.0 & 1 \\
 25  & NGC 3187    &  10 17 48.0  &   +21 52 25  &  1.5 DA  & Apr 09 97  & $r$  &  30  & 13.91 &  5.0 & 1 \\
     &             &              &              &  2.2 ESO & Apr 11 97  & $K'$ &  28  &  ---  &  5.0 & 1 \\
 26  & ESO 317-G29 &  10 27 44.0  & $-$40 26 08  &  0.6 BO  & Dec 28 96  & $R$  &  50  & 13.74 &  1.0 & 1 \\
 27  & ESO 264-G29 &  10 40 12.0  & $-$47 06 11  &  0.6 BO  & Jan 06 96  & $R$  &  60  & 15.68 &  5.6 & 1 \\
     &             &              &              &  2.2 ESO & Apr 12 97  & $K'$ &  11  &  ---  &  5.6 & 1 \\
 28  & NGC 3432    &  10 52 31.0  &   +36 37 08  &  1.1 LO  & May 31 97  & $R$  &  40  & 11.67 &  9.0 & 1 \\
     &             &              &              &  2.2 CA  & Feb 13 98  & $K'$ &  60  &  ---  &  9.0 & 1 \\
 29  & NGC 3628    &  11 20 16.0  &   +13 35 22  &  1.5 DA  & Apr 09 97  & $r$  &  95  & 10.28 &  3.0 & 1 \\
     &             &              &              &  2.2 CA  & Feb 15 98  & $K'$ &  13  &  ---  &  3.0 & 1 \\
 30  & ESO 378-G13 &  11 37 08.0  & $-$32 49 13  &  0.6 BO  & Jan 07 97  & $R$  &  60  & 15.45 &  1.0 & 1 \\
     &             &              &              &  2.2 ESO & Apr 12 97  & $K'$ &  38  &  ---  &  1.0 & 1 \\
 31  & ESO 379-G20 &  12 00 59.0  & $-$35 11 36  &  0.6 BO  & Jan 08 97  & $R$  &  75  & 15.44 &  1.0 & 1 \\
     &             &              &              &  2.2 ESO & Apr 11 97  & $K'$ &  36  &  ---  &  1.0 & 1 \\
%
%
 32  & NGC 4183    &  12 13 18.0  &   +43 41 55  &  1.1 LO  & Jun 02 97  & $R$  &  25  & 12.86 &  6.0 & 1 \\
     &             &              &              &  2.2 CA  & Feb 14 98  & $K'$ &  40  &  ---  &  6.0 & 1 \\
 33  & NGC 4631    &  12 42 08.0  &   +32 32 28  &  1.1 LO  & May 31 97  & $R$  &  30  &  9.75 &  7.0 & 1 \\
     &             &              &              &  2.2 CA  & Feb 16 98  & $K'$ &  13  &  ---  &  7.0 & 1 \\
%
%
  \noalign{\smallskip}
  \hline
  \end{tabular}
  \end{flushleft}
  \end{table*}
\end{center}

\begin{center}
  \begin{table*}
 {\bf Table 4.} continued. \\
  \begin{flushleft}
  \begin{tabular}{rllllcccrrc}
  \cline{1-11}
  \hline\hline
  \noalign{\smallskip}
  \multicolumn{1}{r}{No.}          & \multicolumn{1}{c}{Galaxy}     &
  \multicolumn{1}{c}{RA (2000)}    & \multicolumn{1}{c}{Dec (2000)} &
  \multicolumn{1}{c}{Telescope}    & \multicolumn{1}{c}{Date}       &
  \multicolumn{1}{c}{Band}         & \multicolumn{1}{c}{Exp.-time}  &
  \multicolumn{1}{c}{$B_{\rm T}$}  & \multicolumn{1}{c}{Type}       & \multicolumn{1}{c}{Note}  \\
  \noalign{\smallskip}
   &  & \multicolumn{1}{l}{\hspace*{1mm}h\hspace{3mm}m\hspace{3mm}s}    &
   \multicolumn{1}{c}{\hspace*{1mm}$\degr$\hspace{3mm}$'$\hspace{3mm}$''$}  &  &
   &  & \multicolumn{1}{c}{[min]}  & \multicolumn{1}{c}{[mag]}   &  &  \\
  \noalign{\smallskip}
  \multicolumn{1}{c}{(1)}  &  \multicolumn{1}{c}{(2)}  &  \multicolumn{1}{c}{(3)} &
  \multicolumn{1}{c}{(4)}  &  \multicolumn{1}{c}{(5)}  &  \multicolumn{1}{c}{(6)} &
  \multicolumn{1}{c}{(7)}  &  \multicolumn{1}{c}{(8)}  &  \multicolumn{1}{c}{(9)} &
  \multicolumn{1}{c}{(10)} &  \multicolumn{1}{c}{(11)} \\
  \noalign{\smallskip}
  \hline
  \noalign{\medskip}
%
%
 34  & NGC 4634    &  12 42 40.4  &   +14 17 47  &  1.5 DA  & Jun 03 98  & $r$  &  45  & 13.16 &  6.0 & 1 \\
     &             &              &              &  2.2 CA  & Feb 15 98  & $K'$ &  25  &  ---  &  6.0 & 1 \\
 35  & NGC 4747    &  12 51 45.0  &   +25 46 27  &  1.1 LO  & Jun 02 97  & $R$  &  72  & 12.96 &  6.0 & 1 \\
 36  & NGC 4762    &  12 52 56.3  &   +11 13 48  &  1.5 DA  & Apr 08 97  & $r$  &  30  & 11.12 & -2.0 & 1 \\
     &             &              &              &  2.2 CA  & Feb 15 98  & $K'$ &  27  &  ---  & -2.0 & 1 \\
 37  & ESO 443-G21 &  12 59 46.0  & $-$29 35 58  &  1.5 DA  & Apr 08 97  & $r$  &  45  & 14.41 &  6.0 & 1 \\
     &             &              &              &  2.2 ESO & Apr 11 97  & $K'$ &  15  &  ---  &  6.0 & 1 \\
 38  & NGC 5126    &  13 24 54.0  & $-$30 20 00  &  0.6 BO  & Jan 10 97  & $R$  &  40  & 14.06 &  0.0 & 1 \\
 39  & ESO 324-23  &  13 27 29.0  & $-$38 10 26  &  1.5 DA  & Apr 08 97  & $r$  &  45  & 13.07 &  6.5 & 1 \\
     &             &              &              &  2.2 ESO & Apr 10 97  & $K'$ &  32  &  ---  &  6.5 & 1 \\
 40  & ESO 383-G05 &  13 29 23.8  & $-$34 16 23  &  1.5 DA  & Jun 03 98  & $r$  &  45  & 14.21 &  3.7 & 1 \\
 41  & NGC 5297    &  13 46 24.0  &   +43 52 25  &  1.1 LO  & Jun 03 97  & $R$  &   3  & 12.47 &  4.5 & 1 \\
 42  & ESO 445-G63 &  13 52 07    & $-$30 49 41  &  1.5 DA  & Jun 03 98  & $r$  &  40  & 15.78 &  5.3 & 1 \\
%
%
 43  & NGC 5529    &  14 15 34.1  &   +36 13 36  &  1.1 LO  & Jun 01 97  & $R$  &  50  & 12.75 &  5.0 & 1 \\
     &             &              &              &  1.2 CA  & Mar 05 96  & $H$  &  12  &  ---  &  5.0 & 1 \\
 44  & NGC 5965    &  15 34 02.0  &   +56 41 10  &  1.1 LO  & Jun 03 97  & $R$  &  60  & 12.60 &  3.0 & 1 \\
 45  & NGC 6045    &  16 05 08.0  &   +17 45 22  &  1.2 CA  & Sep 06 96  & $R$  &  30  & 14.87 &  5.0 & 1 \\
     &             &              &              &  2.2 CA  & Sep 03 96  & $K$  &  30  &  ---  &  5.0 & 1 \\
 46  & NGC 6361    &  17 18 40.0  &   +60 36 32  &  1.2 CA  & Jun 02 96  & $R$  &  37  & 13.87 &  3.0 & 1 \\
     &             &              &              &  1.0 HL  & May 20 96  & $nf\A$&  5  &  ---  &  3.0 & 1 \\
     &             &              &              &  2.2 CA  & Sep 03 96  & $K$  &  20  &  ---  &  3.0 & 1 \\
 47  & Arp 121     &  00 59 24.0  & $-$04 48 13  &  1.2 CA  & Sep 04 96  & $R$  &  60  &  ---  &  2.0 & 1 \\
     &             &              &              &  2.2 CA  & Sep 04 96  & $K$  &  16  &  ---  &  2.0 & 1 \\
 48  & ESO 462-G07 &  20 18 23    & $-$27 27 18  &  2.2 ESO & Apr 12 97  & $K'$ &  41  & 15.48 &  4.0 & 1 \\
 49  & IC 4991     &  20 18 23.0  & $-$41 03 01  &  1.5 DA  & Apr 08 98  & $r$  &  30  & 11.56 & -2.0 & 1 \\
%
%
  \noalign{\medskip}
  \hline
  \noalign{\medskip}
  \multicolumn{11}{c}{\bf Non -- Interacting} \\
  \noalign{\medskip}
  \hline
  \noalign{\medskip}
%
%
  1  & UGC 231     &  00 24 02.6  &   +16 29 09  &  1.0 HL  & Aug 19 96  & $R$  &  30  & 13.91 &  6.0 & 1 \\
     &             &              &              &  2.2 CA  & Sep 05 96  & $K$  &  40  &  ---  &  6.0 & 1 \\
  2  & ESO 150-G07 &  00 25 37.0  & $-$57 11 28  &  1.5 DA  & Jun 04 98  & $r$  &  30  & 15.28 &  1.0 & 1 \\
  3  & ESO 112-G04 &  00 28 04.0  & $-$58 06 13  &  2.2 ESO & June 87    & $r$  &      & 15.86 &  5.6 & 2 \\
  4  & ESO 150-G14 &  00 36 38.0  & $-$56 54 24  &  2.2 ESO & June 87    & $r$  &      & 14.90 &  0.4 & 2 \\
  5  & UGC 711     &  01 08 37.0  &   +01 38 29  &  0.6 Bo  & Dec 25 96  & $R$  & 118  & 14.39 &  6.7 & 1 \\
  6  & ESO 244-G48 &  01 39 09.0  & $-$47 07 42  &  1.5 DA  & Jun 03 98  & $r$  &  30  & 15.55 & -2.0 & 1 \\
  7  & UGC 1839    &  02 22 30.2  & $-$00 37 07  &  1.2 CA  & Sep 08 96  & $R$  &  75  & 15.26 &  7.3 & 1 \\
  8  & NGC 891     &  02 22 33.1  &   +42 20 48  &  1.0 HL  & Sep 05 97  & $R$  &  30  & 10.81 &  3.0 & 1 \\
     &             &              &              &  2.2 CA  & Feb 14 98  & $K'$ &  20  &  ---  &  3.0 & 1 \\
  9  & ESO 416-G25 &  02 48 41.0  & $-$31 32 10  &  2.2 ESO & June 87    & $r$  &      & 14.64 &  3.0 & 2 \\
 10  & UGC 2411    &  02 58 00.9  &   +75 45 00  &  1.1 LO  & Mar 17 96  & $R$  &  20  & 16.50 &  8.5 & 1 \\
%
%
 11  & IC 1877     &  03 03 10.0  & $-$50 30 43  &  0.6 BO  & Dec 28 96  & $R$  &  88  & 16.30 &  3.0 & 1 \\
 12  & ESO 201-G22 &  04 09 00.4  & $-$48 43 35  &  0.6 BO  & Dec 25 96  & $R$  & 148  & 14.69 &  5.0 & 1 \\
 13  & NGC 1886    &  05 12 48.7  & $-$23 48 45  &  0.6 BO  & Jan 06 97  & $R$  &  30  & 13.60 &  3.5 & 1 \\
     &             &              &              &  2.2 CA  & Feb 15 98  & $K'$ &  20  &  ---  &  3.5 & 1 \\
 14  & UGC 3474    &  06 32 00.6  &   +71 33 00  &  1.1 LO  & Mar 17 96  & $R$  &  10  & 15.40 &  6.0 & 1 \\
     &             &              &              &  2.2 CA  & Feb 14 98  & $K'$ &  27  &  ---  &  6.0 & 1 \\
 15  & NGC 2310    &  06 53 53.6  & $-$40 51 44  &  0.6 BO  & Jan 05 07  & $R$  &  30  & 12.74 & -2.0 & 1 \\
 16  & UGC 4278    &  08 13 59.0  &   +45 54 43  &  1.1 LO  & Mar 16 96  & $R$  &  10  & 13.07 &  7.0 & 1 \\
     &             &              &              &  1.2 CA  & Mar 02 96  & $H$  &  67  &  ---  &  7.0 & 1 \\
 17  & ESO 564-G27 &  09 11 54.4  & $-$20 07 03  &  2.2 ESO & June 87    & $r$  &      & 14.35 &  6.3 & 2 \\
 18  & UGC 4943    &  09 19 58.1  &   +37 11 27  &  1.1 LO  & Mar 16 96  & $R$  &   6  & 14.80 &  3.0 & 1 \\
     &             &              &              &  1.2 CA  & Mar 06 96  & $H$  &   8  &  ---  &  3.0 & 1 \\
 19  & IC 2469     &  09 23 00.9  & $-$32 26 59  &  1.5 DA  & Jun 02 08  & $r$  &   5  & 13.03 &  2.0 & 1 \\
%
%
 20  & UGC 5341    &  09 56 36.6  &   +20 38 53  &  0.6 BO  & Jan 02 97  & $R$  & 134  & 15.03 &  6.0 & 1 \\
     &             &              &              &  2.2 CA  & Feb 15 98  & $K'$ &  27  &  ---  &  6.0 & 1 \\
%
%
  \noalign{\smallskip}
  \hline
  \end{tabular}
  \end{flushleft}
  \end{table*}
\end{center}

\begin{center}
  \begin{table*}
 {\bf Table 4.} continued. \\
  \begin{flushleft}
  \begin{tabular}{rllllcccrrc}
  \cline{1-11}
  \hline\hline
  \noalign{\smallskip}
  \multicolumn{1}{r}{No.}          & \multicolumn{1}{c}{Galaxy}     &
  \multicolumn{1}{c}{RA (2000)}    & \multicolumn{1}{c}{Dec (2000)} &
  \multicolumn{1}{c}{Telescope}    & \multicolumn{1}{c}{Date}       &
  \multicolumn{1}{c}{Band}         & \multicolumn{1}{c}{Exp.-time}  &
  \multicolumn{1}{c}{$B_{\rm T}$}  & \multicolumn{1}{c}{Type}       & \multicolumn{1}{c}{Note}  \\
  \noalign{\smallskip}
   &  & \multicolumn{1}{l}{\hspace*{1mm}h\hspace{3mm}m\hspace{3mm}s}    &
   \multicolumn{1}{c}{\hspace*{1mm}$\degr$\hspace{3mm}$'$\hspace{3mm}$''$}  &  &
   &  & \multicolumn{1}{c}{[min]}  & \multicolumn{1}{c}{[mag]}   &  &  \\
  \noalign{\smallskip}
  \multicolumn{1}{c}{(1)}  &  \multicolumn{1}{c}{(2)}  &  \multicolumn{1}{c}{(3)} &
  \multicolumn{1}{c}{(4)}  &  \multicolumn{1}{c}{(5)}  &  \multicolumn{1}{c}{(6)} &
  \multicolumn{1}{c}{(7)}  &  \multicolumn{1}{c}{(8)}  &  \multicolumn{1}{c}{(9)} &
  \multicolumn{1}{c}{(10)} &  \multicolumn{1}{c}{(11)} \\
  \noalign{\smallskip}
  \hline\noalign{\medskip}
%
%
 21  & IC 2531     &  09 59 55.7  & $-$29 36 55  &  1.5 DA  & Apr 08 97  & $r$  &  30  & 12.90 &  5.3 & 1 \\
 22  & NGC 3390    &  10 48 04.0  & $-$31 31 57  &  1.5 DA  & Jun 02 98  & $r$  &  30  & 12.85 &  3.0 & 1 \\
 23  & ESO 319-G26 &  11 30 20.0  & $-$41 03 57  &  2.2 ESO & June 87    & $r$  &      & 14.58 &  5.3 & 2 \\
 24  & NGC 3957    &  11 54 01.1  & $-$19 34 06  &  1.5 DA  & Jun 03 98  & $r$  &  15  & 12.81 & -1.0 & 1 \\
 25  & NGC 4013    &  11 58 31.7  &   +43 56 48  &  1.1 LO  & Jun 01 97  & $R$  &  18  & 12.19 &  3.0 & 1 \\
     &             &              &              &  2.2 CA  & Feb 15 98  & $K'$ &  33  &  ---  &  3.0 & 1 \\
 26  & ESO 572-G44 &  12 01 09.0  & $-$20 29 18  &  1.5 DA  & Jun 03 98  & $r$  &  20  & 15.04 &  3.0 & 1 \\
 27  & UGC 7170    &  12 10 37.0  &   +18 49 24  &  1.1 LO  & Mar 17 96  & $R$  &  15  & 14.96 &  6.0 & 1 \\
 28  & ESO 321-G10 &  12 11 42.0  & $-$38 32 53  &  2.2 ESO & June 87    & $r$  &      & 14.22 &  1.4 & 2 \\
%
%
 29  & NGC 4217    &  12 15 50.9  &   +47 05 32  &  1.1 LO  & Jun 01 97  & $R$  &  30  & 12.04 &  3.0 & 1 \\
 30  & NGC 4244    &  12 17 30.0  &   +37 48 27  &  1.1 LO  & Jun 01 97  & $R$  &  20  & 10.88 &  6.0 & 1 \\
 31  & UGC 7321    &  12 17 34.1  &   +22 32 21  &  1.1 LO  & Mar 16 96  & $R$  &  20  & 14.15 &  7.0 & 1 \\
     &             &              &              &  1.2 CA  & Mar 02 96  & $H$  &  59  &  ---  &  7.0 & 1 \\
 32  & NGC 4302    &  12 21 42.5  &   +14 36 05  &  1.0 HL  & May 19 96  & $R$  &  10  & 12.50 &  5.0 & 1 \\
     &             &              &              &  1.2 CA  & Mar 02 96  & $H$  &  53  &  ---  &  5.0 & 1 \\
 33  & NGC 4330    &  12 23 16.5  &   +11 22 06  &  1.5 DA  & Apr 09 97  & $r$  &  40  & 13.09 &  6.0 & 1 \\
     &             &              &              &  2.2 ESO & Apr 10 97  & $K'$ &  32  &  ---  &  6.0 & 1 \\
 34  & NGC 4565    &  12 36 20.6  &   +25 59 05  &  1.0 HL  & May 20 96  & $R$  &  10  & 10.42 &  3.0 & 1 \\
     &             &              &              &  2.2 CA  & Feb 16 98  & $K'$ &  13  &  ---  &  3.0 & 1 \\
 35  & NGC 4710    &  12 49 39.0  &   +15 09 55  &  1.1 LO  & Jun 03 97  & $R$  &  15  & 11.91 & -1.0 & 1 \\
     &             &              &              &  2.2 CA  & Feb 16 98  & $K'$ &  20  &  ---  & -1.0 & 1 \\
%
%
 36  & NGC 5170    &  13 29 49.0  & $-$17 57 59  &  1.5 DA  & Apr 08 97  & $r$  &  45  & 12.06 &  5.0 & 1 \\
     &             &              &              &  1.2 CA  & Mar 03 96  & $H$  &  11  &  ---  &  5.0 & 1 \\
     &             &              &              &  1.2 CA  & Mar 05 96  & $K$  &   8  &  ---  &  5.0 & 1 \\
 37  & ESO 510-G18 &  13 55 32.0  & $-$27 24 47  &  1.5 DA  & Jun 03 98  & $r$  &  30  & 16.21 &  1.0 & 1 \\
 38  & UGC 9242    &  14 25 20.9  &   +39 32 22  &  1.1 LO  & Mar 15 96  & $R$  &  21  & 14.09 &  7.0 & 1 \\
     &             &              &              &  1.2 CA  & Mar 04 96  & $H$  &  60  &  ---  &  7.0 & 1 \\
 39  & NGC 5775    &  14 53 57.7  &   +03 32 40  &  1.5 DA  & Jun 03 98  & $r$  &  15  & 12.24 &  5.0 & 1 \\
 40  & NGC 5907    &  15 15 53.8  &   +56 19 46  &  1.2 CA  & Jun 02 96  & $R$  &  30  & 11.12 &  5.0 & 1 \\
     &             &              &              &  1.2 CA  & Jun 04 96  & $H$  &   6  &  ---  &  5.0 & 1 \\
 41  & NGC 5908    &  15 16 43.5  &   +55 24 40  &  1.2 CA  & Sep 08 96  & $R$  &  45  & 12.79 &  3.0 & 1 \\
     &             &              &              &  2.2 CA  & Feb 16 98  & $K'$ &  33  &  ---  &  3.0 & 1 \\
 42  & ESO 583-G08 &  15 57 50.5  & $-$22 29 47  &  1.5 DA  & Jun 04 98  & $r$  &  35  &  ---  &  4.0 & 1 \\
 43  & NGC 6181    &  16 32 20.9  &   +19 49 30  &  1.1 LO  & Jun 03 97  & $R$  &  15  & 12.49 &  5.0 & 1 \\
 44  & ESO 230-G11 &  18 46 24.0  & $-$52 09 23  &  1.5 DA  & Jun 04 98  & $r$  &  30  & 13.74 &  4.0 & 1 \\
 45  & NGC 6722    &  19 03 40.0  & $-$64 53 41  &  2.2 ESO & June 87    & $r$  &      & 13.54 &  3.0 & 2 \\
 46  & ESO 461-G06 &  19 51 55.9  & $-$31 58 52  &  1.5 DA  & Jun 04 98  & $r$  &  40  & 16.21 &  5.0 & 1 \\
%
%
 47  & ESO 339-G16 &  20 00 07.0  & $-$40 43 03  &  2.2 ESO & June 87    & $r$  &      & 16.50 &  1.0 & 2 \\
 48  & IC 4937     &  20 05 18.0  & $-$56 15 20  &  2.2 ESO & June 87    & $r$  &      & 14.86 &  3.0 & 2 \\
 49  & ESO 187-G08 &  20 43 25.2  & $-$56 12 17  &  1.5 DA  & Jun 04 98  & $r$  &  30  & 15.69 &  6.0 & 1 \\
 50  & IC 5052     &  20 52 06.3  & $-$69 12 14  &  1.5 DA  & Jun 03 98  & $r$  &  35  & 11.16 &  7.0 & 1 \\
     &             &              &              &  2.2 ESO & Apr 10 97  & $K'$ &  31  &  ---  &  7.0 & 1 \\
 51  & IC 5096     &  21 18 22.0  & $-$63 45 41  &  1.5 DA  & Jun 03 98  & $r$  &  30  & 13.30 &  4.0 & 1 \\
 52  & ESO 466-G01 &  21 42 32.0  & $-$29 22 10  &  2.2 ESO & June 87    & $r$  &      & 14.63 &  2.0 & 2 \\
 53  & ESO 189-G12 &  21 55 38.7  & $-$54 52 33  &  1.5 DA  & Jun 04 98  & $r$  &  30  & 15.59 &  5.0 & 1 \\
 54  & UGC 11859   &  21 58 07.3  &   +01 00 34  &  2.2 ESO & June 87    & $r$  &      & 15.16 &  4.0 & 2 \\
 55  & ESO 533-G04 &  22 14 03.2  & $-$26 56 18  &  1.5 DA  & Jun 03 98  & $r$  &  70  & 14.18 &  4.8 & 1 \\
 56  & IC 5199     &  22 19 33.0  & $-$37 32 01  &  1.5 DA  & Jun 04 98  & $r$  &  30  & 15.00 &  3.0 & 1 \\
%
%
 57  & UGC 11994   &  22 20 53.4  &   +33 17 34  &  1.0 HL  & Sep 07 97  & $R$  &  54  & 14.85 &  4.0 & 1 \\
 58  & UGC 12281   &  22 59 12.4  &   +13 36 21  &  1.2 CA  & Sep 07 96  & $R$  &  60  & 14.79 &  8.0 & 1 \\
     &             &              &              &  2.2 CA  & Sep 04 96  & $K$  &  20  &  ---  &  8.0 & 1 \\
 59  & UGC 12423   &  23 13 06.0  &   +06 24 00  &  1.2 CA  & Sep 07 96  & $R$  &  60  & 14.53 &  5.0 & 1 \\
     &             &              &              &  2.2 CA  & Sep 03 96  & $K$  &  27  &  ---  &  5.0 & 1 \\
 60  & NGC 7518    &  23 13 12.9  &   +06 19 16  &  1.1 LO  & Jun 02 97  & $R$  &  40  & 14.24 &  1.0 & 1 \\
 61  & ESO 604-G06 &  23 14 54.0  & $-$20 59 44  &  1.5 DA  & Jun 03 98  & $r$  &  40  & 15.00 &  4.0 & 1 \\
%
%
  \noalign{\smallskip}
  \hline
  \end{tabular}
  \begin{list}{}{}
  \item[$\A$] nf= no filter used. \\
  \end{list}
  \end{flushleft}
  \end{table*}
\end{center}

\clearpage






\begin{figure*}[t]
\vspace*{108mm}
\hspace*{8mm}
\begin{minipage}[b]{5.5cm}
\begin{picture}(4.0,10.5)
{\includegraphics[angle=90,viewport=36 223 530 420,clip,width=44mm]{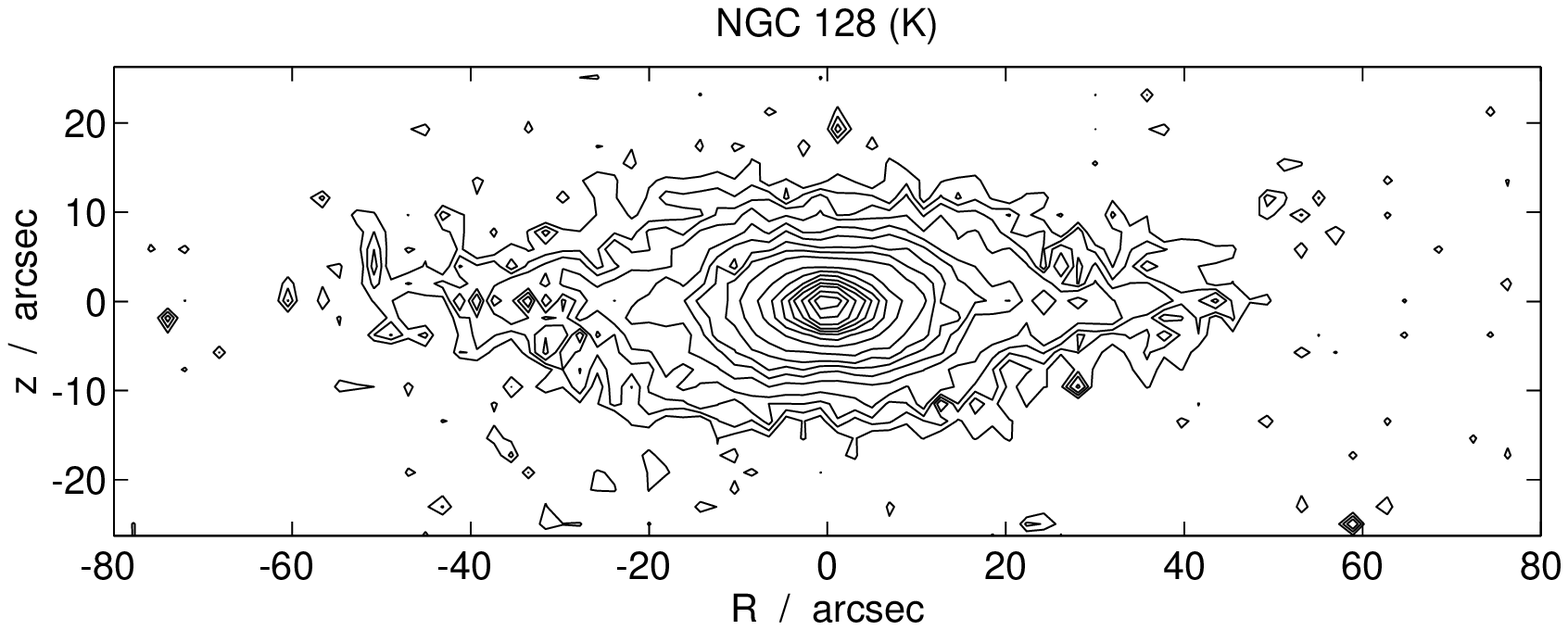}}
\end{picture}
\end{minipage}
\hfill
\begin{minipage}[b]{5.5cm}
\begin{picture}(4.0,11)
{\includegraphics[angle=90,viewport=36 223 530 420,clip,width=44mm]{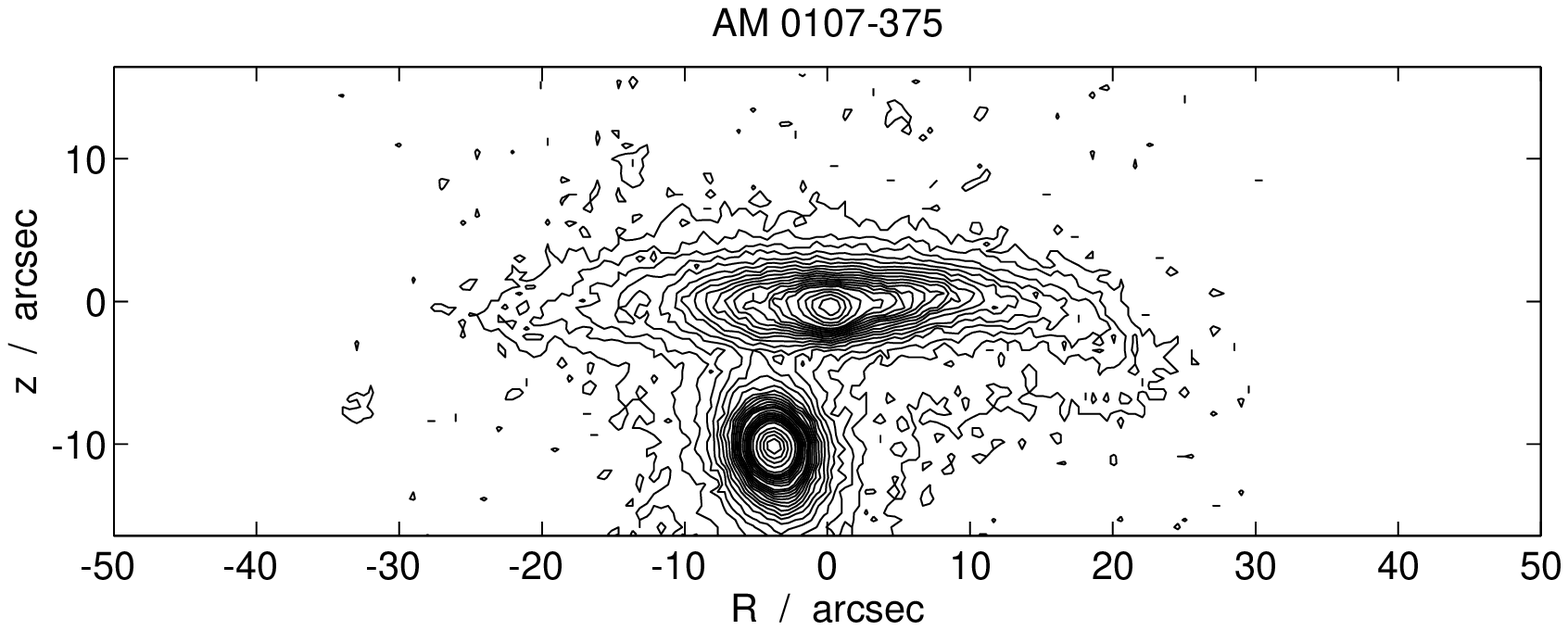}}
\end{picture}
\end{minipage}
\hfill
\begin{minipage}[b]{5.5cm}
\begin{picture}(4.0,11)
{\includegraphics[angle=90,viewport=36 223 530 420,clip,width=44mm]{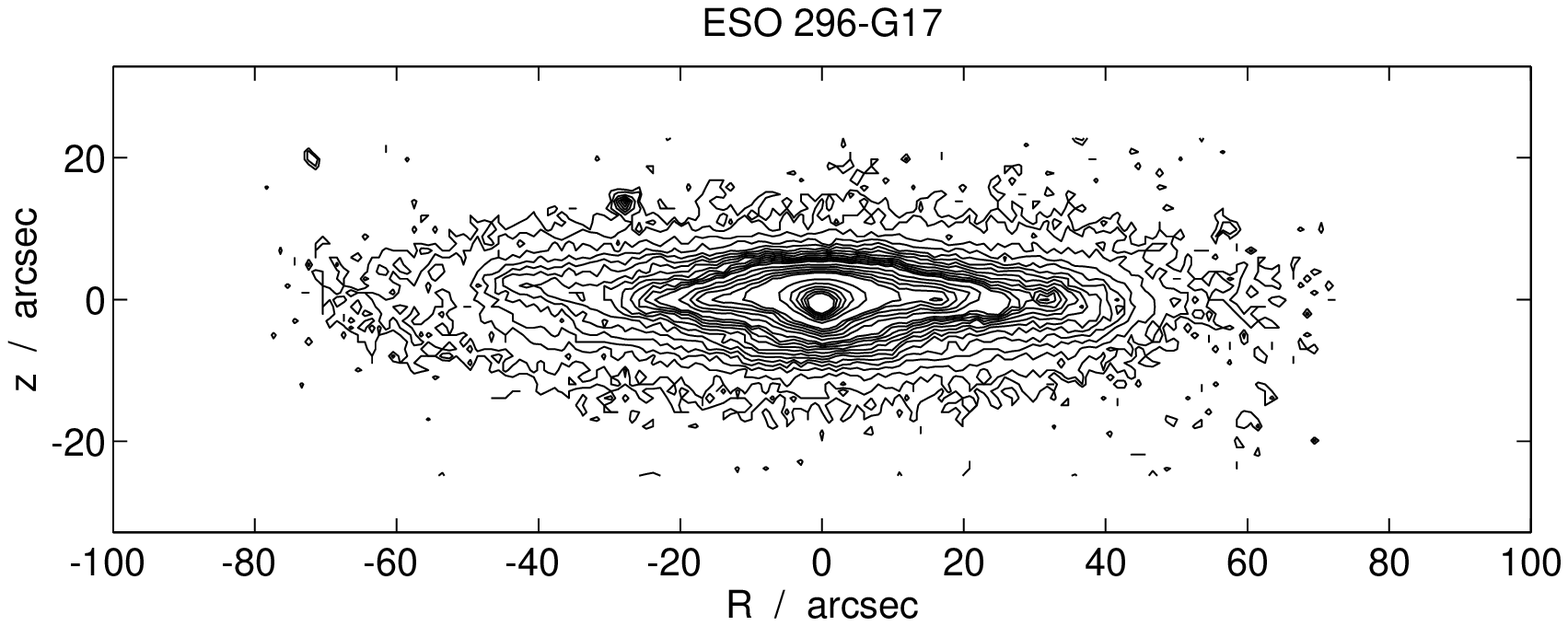}}
\end{picture}
\end{minipage}
\\

\vspace*{104mm}
\hspace*{8mm}
\begin{minipage}[b]{5.5cm}
\begin{picture}(4.0,11)
{\includegraphics[angle=90,viewport=36 223 530 420,clip,width=44mm]{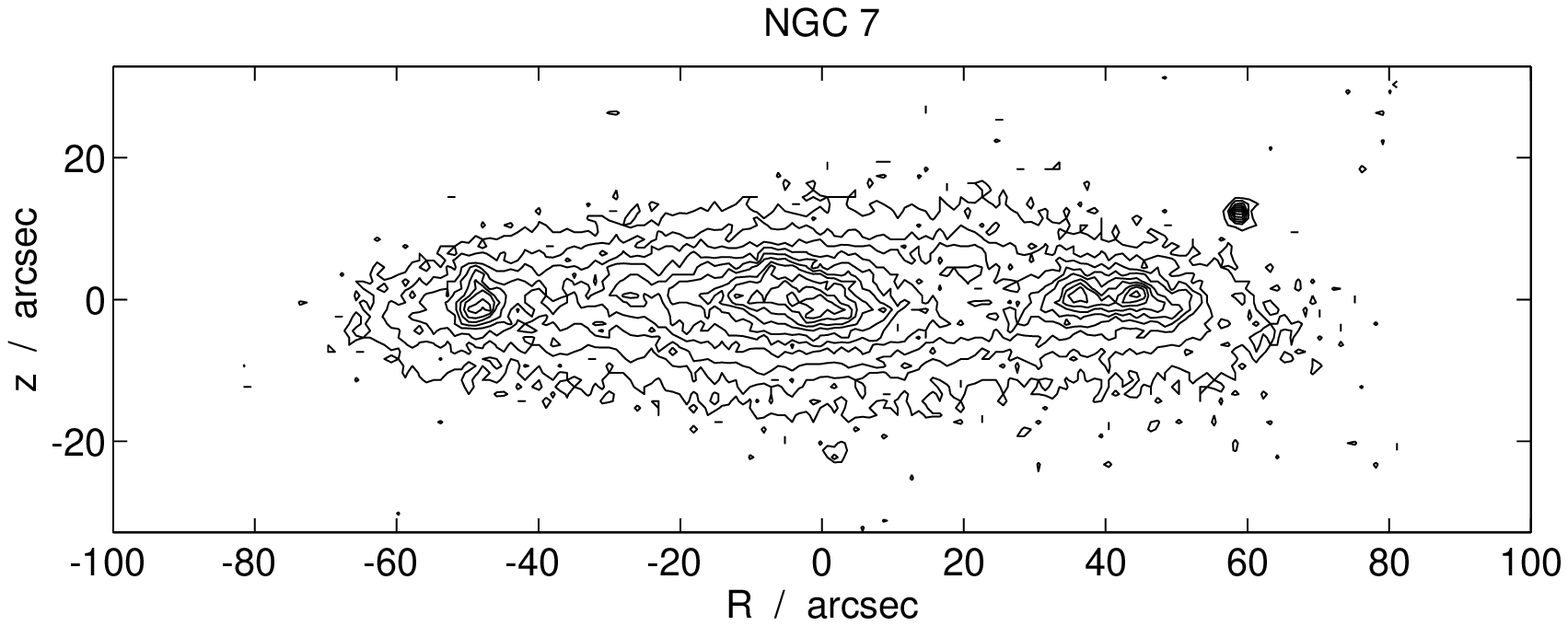}}
\end{picture}
\end{minipage}
\hfill
\begin{minipage}[b]{5.5cm}
\begin{picture}(4.0,11)
{\includegraphics[angle=90,viewport=36 223 530 420,clip,width=44mm]{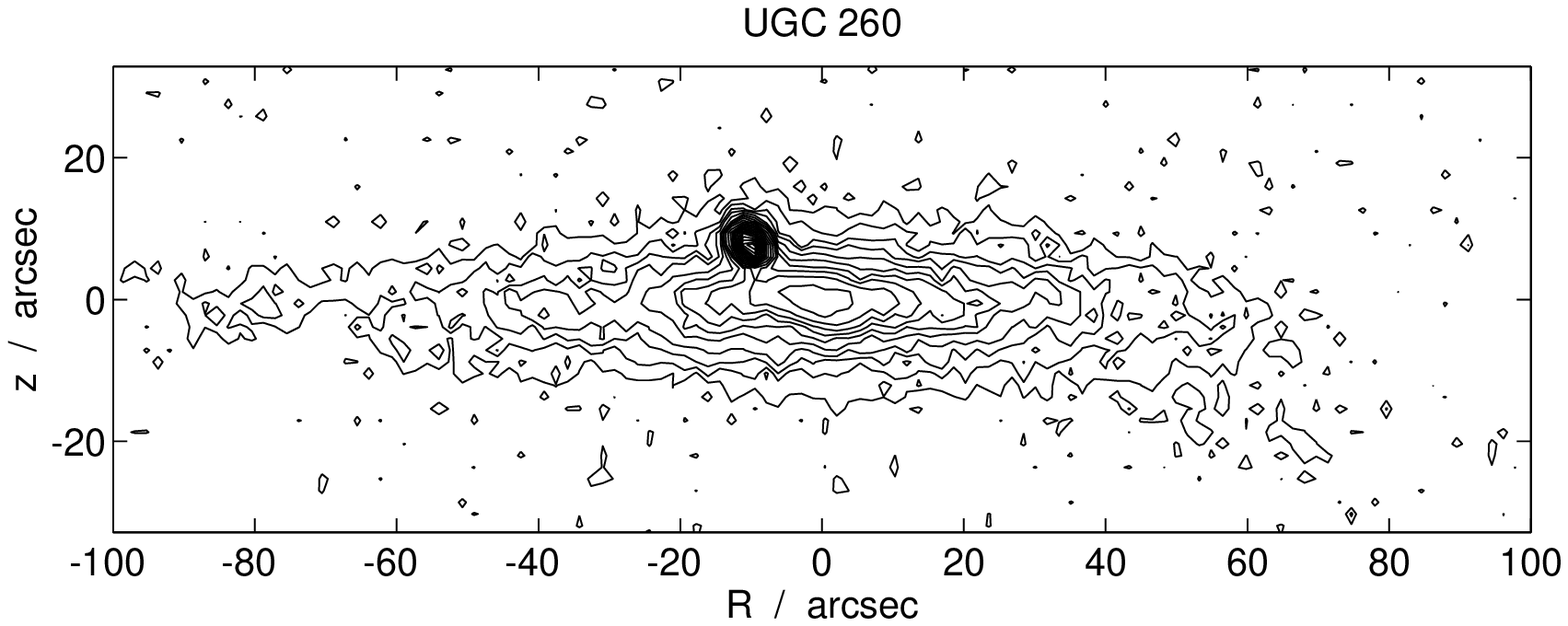}}
\end{picture}
\end{minipage}
\hfill
\begin{minipage}[b]{5.5cm}
\begin{picture}(4.0,11)
{\includegraphics[angle=90,viewport=36 223 530 420,clip,width=44mm]{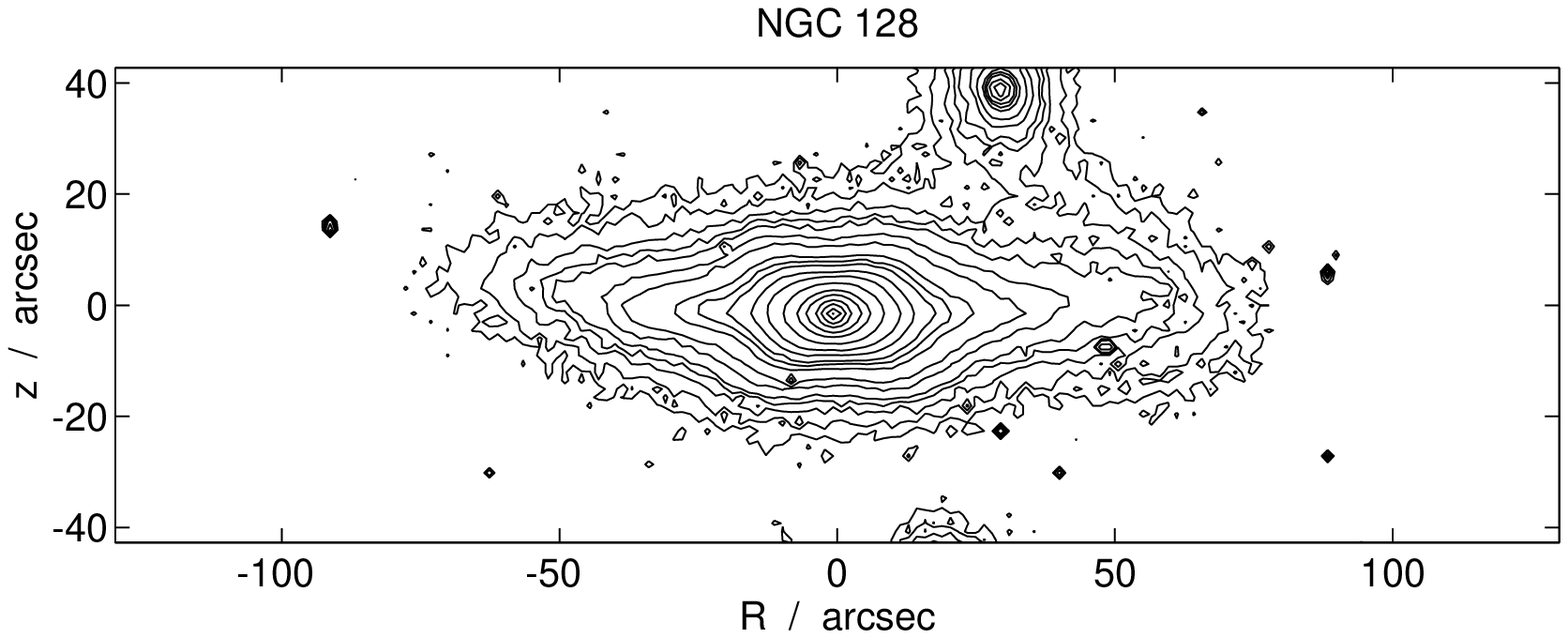}}
\end{picture}
\end{minipage}
\\ \\
\hspace*{14mm}\parbox{150mm}{
{\bf \noindent Fig. 4.} Optical ($R$, $r$) and near infrared ($H$, $K'$, $K$) contour maps
of the sample of interacting/merging galaxies studied in this paper (see Table~\ref{samples}).
The contour interval is 0.4 mag, starting at a faintest level of 0.5 mag above the noise level
of the corresponding image.
}
\end{figure*}



\end{document}